\documentclass[10pt]{article}

\usepackage{amsfonts}
\def\form#1{(\ref{#1})}
\def\Co{I \kern-.66em C}
\def\l{\lambda}

\def\r{\rho}
\def\s{\sigma}
\def\g{\gamma}
\def\al{\alpha}
\def\b{\beta}

\def\o{\omega}
\def\pa{\partial}

\def\L{{\mathcal L}}               %densita` lagrangiana
 \def\U{{\cal U}}                           %Lagrang. duale
 \def\B{{\cal B}} 
\def\H{{\mathcal H}}               %densita` hamiltoniana

\def\E{{\cal E}}  
\def\W{{\cal W}}  
\def\U{{\mathcal U}}               %lagrangiana U

\def\ram{\mathop{\longrightarrow}\limits}
\def\um{\mathop{=}\limits}

\def\dim{\mathop{\rm dim}\nolimits}

\def\Ga#1#2#3{\Gamma^{#1}{}_{#2#3}}

\renewcommand{\Re}{I\kern-.36em R}         %R dei reali 

\newcommand{\be}{\begin{equation}}
\newcommand{\ee}{\end{equation}}
\newcommand{\ba}{\begin{eqnarray}}
\newcommand{\ea}{\end{eqnarray}}
\newcommand{\baa}{\be\left\{\begin{array}{l}}
\newcommand{\eaa}{\end{array}\right.\ee}

\def\QDE{\rule{2.5mm}{2.5mm}}
\def\CVD{$\phantom{'}$\hfill\QDE}

\newtheorem{Theorem}{Theorem}[section]

\newtheorem{Remark}[Theorem]{Remark} %
\newtheorem{Definition}[Theorem]{Definition} %
\newtheorem{Lemma}[Theorem]{Lemma} %
\newtheorem{Example}[Theorem]{Example} %
\newtheorem{Proposition}[Theorem]{Proposition} %
\newtheorem{Exercise}[Theorem]{Exercise}%

\title{Hamiltonian, Energy and Entropy in
General Relativity with Non--Orthogonal Boundaries}
\author{M.\ Francaviglia\thanks{E-mail:
francaviglia@dm.unito.it}, M.\ Raiteri\thanks{E-mail:
raiteri@dm.unito.it}
\\
Dipartimento di Matematica, Universit\`a degli
Studi di Torino,\\
Via Carlo Alberto 10, 10123 Torino, Italy }

\date{}

\begin{document}
\maketitle

\begin{abstract}
A general recipe to define, via N\"other theorem, the Hamiltonian in
any natural field theory is suggested. It is based on a
Regge--Teitelboim--like approach applied to the variation of N\"other 
conserved quantities. The Hamiltonian for General Relativity in
presence of non--orthogonal boundaries  is analysed  
and the energy  is defined  as the on--shell value of the
Hamiltonian. The role played   by boundary conditions in the formalism
is outlined and the quasilocal internal energy  is defined by
imposing metric Dirichlet boundary conditions. A (conditioned)
agreement with previous definitions is proved. A 
correspondence with Brown-York original formulation of the first
principle  of black hole thermodynamics is  finally established.

\end{abstract}
%%%%%%%%%%%%%%%%%%%%%%%%%%%%%%%%%%%%%%%%%%%%%%%%%%%%%%%%%%%%%%%%%%%%%%%%%%%%%%%

\section{Introduction}

One of the most popular approaches  to black hole thermodynamics  is
due to Brown and York; see \cite{BY,BYdue}. They defined,  through an
Hamilton--Jacobi  analysis of a gravitational system,  the quasilocal
energy, the  quasilocal angular momentum and the spatial stress of a
spatially bounded  region of spacetime.
Those quasilocal surface densities were  subsequently involved  in
describing  the first law of black hole thermodynamics through a
statistical analysis  based  on path integral methods. As a result,
in the final expression of the first law, quasilocal physical
quantities  are involved rather than the global ones calculated
asymptotically.
 Some advantages may, in fact,  be gained by
considering  spatially bounded 
self--gravitating systems  instead of  considering the properties of
the whole universe. 
 From a
mathematical viewpoint,  if we analyze only a spatially bounded region,
the asymptotic behavior of the
gravitational field becomes  irrelevant and different gravitational
solutions  with  different asymptotic behaviors can be handled on
equal footing. On the other hand, from a physical viewpoint
it was noticed that a system in thermal equilibrium must feature  a
finite spatial extent: a system of infinite spatial extent  at fixed
temperature  is thermodynamically unstable (see e.g.
\cite{BTZRefB}). Hence, one could argue that, in establishing  the
first principle of black hole
thermodynamics,
the
role  of  internal energy
   should be played  by a suitable defined  ``surface  energy'' rather
than by the total mass (calculated in asymptotic regions).
   Correspondingly, if   the temperature is identified with the
intensive variable which is
thermodynamically conjugated  to the energy, it has to depend somehow 
on the surface rather than
to be identified with a constant parameter.
The same reasoning can be applied to all the relevant physical
quantities which are expected to appear in the first law of black hole
thermodynamics.
This is the idea at the basis of the
aforementioned papers
\cite{BY}, \cite{BYdue}.

Nevertheless the formalism developed  in those papers deals only with
foliations of spacetime  with spatial boundaries  orthogonal to the
timelike boundary.

In \cite{Booth,BLY} the assumption of orthogonal boundaries  was
removed  and the quasilocal formalism of Brown--York  was extended in
order to deal with the general case. 
The formalism there developed  starts  from the so--called Trace-K 
gravitational action functional 
 as introduced  by Hayward in \cite{Hayward}. Roughly speaking, the
Trace--K action differs from the standard Hilbert action by metric
dependent boundary and corner terms  and it features a variational
principle with  fixed boundary metric.
The canonical analysis of \cite{Booth,BLY} goes on   through an ADM
splitting  of the Trace-K action functional. The Hamiltonian 
ensues from  the canonical reduction of the action; the energy
(or it would be better to say the mass) is defined as the on--shell
value of the Hamiltonian and  turns out to be a pure boundary term. By
focusing  just on  the (space $+$ time) foliation of the
timelike boundary, the formalism turns out to be  independent  on the
foliation of   spacetime inside or outside  the boundary.  The
expression of the quasilocal energy for foliations with 
non--orthogonal boundaries was  applied  in
\cite{Booth} to study  the Lorentz--like transformation rules of the
quasilocal energy with respect to boosted observers. In particular, 
the energy measured  by observers infalling towards a black hole was
calculated. Moreover, the close analogy between  the quasilocal
formalism and the thin shell  formalism of Israel (see
\cite{Gravitation}) was analysed in \cite{Booth} with the aim  of
physically supporting and motivating  the  interpretation of energy
attributed to the  mathematical formula for energy there suggested.

A similar formulation was also given by Hawking and Hunter in
\cite{HHnon}. However the analysis there developed was carried on  by
considering  the foliation of the spacetime as a whole  rather than
focusing just on  the foliation of the boundary. The Hawking--Hunter
definition of the gravitational Hamiltonian 
features  a ``tilting term''  which has the disadvantage  to depend
on the intersection angle of the boundaries. In order  to remove  this
unaccettable term a careful choice of the background  has to be
imposed; see \cite{HHnon}.

Another definition of energy for non--orthogonal boundaries can be
also found in  Kijowski's paper \cite{Kij}. Such a definition  was
there obtained by means of a symplectic analysis of the gravitational
Lagrangian and through boundary Legendre transformations. The main
contribution  of Kijowski's approach is related to the way in
which different boundary conditions are related to different
definitions of energy (i.e. internal energy or free energy). Starting
from the same gravitational Lagrangian, namely the Hilbert Lagrangian,
different definitions of energy are recovered in \cite{Kij} through
different control modes of the boundary data. Moreover,
 different boundary conditions reverberate in different
Legendre transformations or, in other words, in different boundary
symplectic structures.

Many attempts to describe  the quasilocal stress--energy content  of
a spatially bounded region of spacetime via N\"other theorem can be
also found elsewhere; see, e.g. \cite{Silva,Wald,Wald95,BYOur} and
references quoted therein. Nevertheless it was long but
wrongly believed -- see
\cite{BLY} -- that a N\"other framework  for conserved quantities 
is not well--suited  to deal with foliations of spacetime with
non--orthogonal boundaries. The purpose of this paper  is to
contradict this common belief. Indeed, we shall show that a
Regge--Teitelboim--like approach to N\"other theorem allows  to
handle boundary terms in such a way that  quasilocal N\"other charges
can be defined. Roughly speaking, the leading idea is to define
\emph{the variation of conserved quantities} $\delta Q$  along  a
$1$--parameter family of field configurations rather than   defining 
the conserved quantities $Q$ in their own. It is in fact well--known
that
\emph{absolute} conserved quantities do not have a proper physical
meaning
while it is instead  meaningful to define
\emph{relative} conserved quantities, i.e. conserved quantities with
respect to a fixed solution chosen as a background and which is
introduced  with the aim to define the ``zero level'' for conserved
quantities themselves; see \cite{Silva,Remarks,Nester,
PetrovKatz,Cavalese}. The background solution has to be chosen
carefully (and differently case by case) because it has to satisfy
appropriate boundary matching conditions. However it is also
well--known that the problem of finding  a suitable background, if
possible,   can be rather cumbersome in direct applications.
Nevertheless, at a theoretical level, the problem can be bypassed by
defining the variation $\delta Q$ as suggested above.
We remind that to each infinitesimal Lagrangian symmetry
$\xi$ one can associate via N\"other theorem a conserved quantity
$Q(\xi)$ which is made of  a bulk term $Q_{_{\hbox{\scriptsize
bulk}}}(\xi)$  plus a surface term $Q_{{\hbox{\scriptsize
surface}}}(\xi)$. On a spacetime
$M$ of arbitrary dimension
$m$ ($m\neq 2$) the bulk term is obtained  by integrating on a 
$(m-1)$ dimensional region $\Sigma$  a suitable
$(m-1)$--form $\tilde \E(\xi)$ which is called
the
\emph{reduced N\"other current}  and which is vanishing on shell,
i.e. by setting
$Q_{_{\hbox{\scriptsize bulk}}}(\xi)=\int_{\Sigma} \tilde \E(\xi)$. 
The surface term
$Q_{{\hbox{\scriptsize surface}}}(\xi)$
 is instead obtained by integrating on the boundary $\partial
\Sigma$  the $(m-2)$--form $\U(\xi)$
which is called
the
\emph{superpotential} and satisfies a canonical splitting
$\E=\tilde\E + d\, \U$, where $\E$ is the generator of N\"other
currents (see e.g. \cite{Robutti,Lagrange}), i.e. by setting
$Q_{{\hbox{\scriptsize surface}}}(\xi)=\int_{\partial
\Sigma}\U(\xi)$.
The superpotential is algorithmically
calculated for every natural or gauge--natural theory starting from
the Lagrangian of the theory and the vector field $\xi$; see
\cite{Remarks,Robutti}. If  we  consider a Cauchy hypersurface
$\Sigma$ in spacetime and a vector field
$\xi$ transverse to $\Sigma$ the  Hamiltonian 
 can be a priori
defined for a Lagrangian
theory as the N\"other charge relative to $\xi$, i.e.
$H(\xi)=Q(\xi)$. 
Namely, the Hamiltonian  is the N\"other charge associated to the
``time'' desplacement  vector field  $\xi$, having identified the
(local)  flow parameter  of  $\xi$ with the (local)
time.
 Nevertheless this definition does not produce the expected
values when specialized to specific solutions (it is indeed
affected by anomalous factors or divergence problems) mainly because
of the fact that such a definition does not take into  proper
account a reference background. Therefore, this prescription  leads
to a definition  of conserved quantities  which has to be somehow
corrected.
 To this end one defines the 
variation $\delta \hat Q$ of the \emph{corrected} conserved
quantity
$\hat Q$ by adding to
$\delta  Q$ a suitable additional boundary term. The latter ensues
from the variation
$\delta Q_{_{\hbox{\scriptsize bulk}}}$ of the bulk term and it keeps
into account the behavior of the fields  at the boundary. In this way
the formalism resembles the original idea of
Regge--Teilteiboim (see \cite{RT}), i.e. boundary terms arising in the
variation of the bulk Hamiltonian are added (with a minus sign) to the
very definition  of Hamiltonian so that the  variation of the
new Hamiltonian  does not  contain boundary terms at all. However,
once  the variation $\delta \hat Q$ of the corrected conserved
quantity is defined,  it remains to analyse the problem if
the variation $\delta \hat Q$ can be formally integrated.
We shall see that this problem is tightly related with the boundary
conditions we choose (i.e. it is related to the control boundary
data). When
$\delta
\hat Q$ is integrable we obtain the conserved quantity $\hat Q-\hat
Q_0$ defined up to a constant of integration $\hat Q_0$. The latter 
can be fixed as a zero level for the conserved quantity or, in other
words,  as a background reference. Namely, $\hat Q_0$
corresponds to the N\"other charge relative  to a solution inside the
set of solutions satisfying the same boundary (or
asymptotic) conditions (and, in turn,  it fixes the boundary or
asymptotic conditions themselves).
  Notice, however, that  the suitable matching conditions between a
solution and its background are required  only on the boundary  (or
at infinity) so that the background is allowed to have a different
topology (for example, this is the case if we fix the Minkowski
background for the Schwarzschild solution; see \cite{BYOur} for
details).

The
variation of N\"other conserved quantities was already  considered
in  \cite{CADM}.  This approach 
  was at the basis of the so--called 
\emph{covariant ADM formalism}. The formalism has been deeply
analysed  and tested in its applications to General Relativity  and it
was shown that it reproduces  the expected physical values of
conserved quantities. Nevertheless, as far as we know,  all the
examples so far analysed in General Relativity cover  only the
non--orthogonal boundary case. It is our goal in this paper to
generalise and extend the aforementioned formalism in order to fill
this gap, also in view of a unifying perspective on the various
formalisms recalled above  and sometimes apparently unrelated.
  To this end  a new additional surface  term has to be
added to the N\"other--based definition of Hamiltonian 
 when
dealing with  non orthogonal boundaries. Indeed the variation of the
corrected  Hamiltonian $\hat H(\xi)=\hat Q(\xi)$ gives rise to an
additional boundary term  which is instead vanishing in presence of
orthogonal intersections. Again,  this term has to be added to $\hat
H$ with a minus sign so that, eventually, the variation of the
Hamiltonian does not really contain surface terms. It is  also
remarkable  that  the final formula we shall obtain  for the
Hamiltonian is independent on divergence terms added to the
Lagrangian, so that  it depends only on the field equations  content 
of the equivalence class $[L]$  of all such dynamically equivalent
Lagrangians.
From a mathematical viewpoint this is clearly an
advantage. Indeed, we do not have to care about surface terms in the
action functional. On the contrary, other Hamiltonian definitions
based on a  framework different from the N\"other one (such as the
ones  based on the canonical reduction of the action functional or
based on a Hamilton--Jacobi analysis) are deeply  sensitive to 
boundary terms. In those formulations, in fact,  boundary terms
reflect into  boundary conditions which have to be satisfied by the
dynamical fields. Different surface terms in the action functional 
then correspond to different boundary conditions and, in the end, to
different definitions of energy. 
Indeed, it is important to notice that for most  physical systems
there may exist   different kinds of energy, each one
corresponding  to different choiches of boundary conditions  and/or of
control variables.  For istance, if we agree that  the internal
energy of a system is the energy corresponding  to Dirichlet boundary
conditions, when we want to calculate  the internal energy for a  
solution of Lagrangian field equations,  we have  to go back to the 
Lagrangian itself and to look for divergence  terms which have to be
added to the Lagrangian to obtain the appropriate
  Dirichlet
boundary conditions (in  General Relativity  the sought--for
Lagrangian  corresponds to the Hayward Lagrangian \cite{Hayward}). 
Moreover, when we try to consider a self--gravitating system  as a
thermodynamical system, different boundary conditions  reverberate 
on different choices  between the fields  which have to be
considered  as intensive or extensive  thermodynamical variables; see
\cite{BYdue}. Accordingly to the choice made on thermodynamical
variables (i.e. on the boundary control modes of the system) one has
then to exibit  different action functionals, e.g.  the
microcanonical action functional, the canonical partition function 
or the grand canonical 
 partition function.
In this way the initial Lagrangian must be modified
each time according to the model one is dealing with.

On the other hand, the definition of  the
Hamiltonian based on N\"other theorem is unique inside each class
$[L]$ of Lagrangians, where the elements of each class  $[L]$ differ
from each other only by  total divergences\footnote{
This is a cohomology class  in the proper  variational cohomology
theory; see, e.g. \cite{MM,Kolar}.}. 
No \emph{ad hoc}
physical or mathematical  prescription is then required  to select a
well--defined representative inside $[L]$ because all of them lead
to the same Hamiltonian.   Nevertheless, if the Hamiltonian is unique
for any element of the class, the question can arise how different
definitions of energy can be recovered  in this framework. In other
words, how different boundary conditions can be imposed on the same
Hamiltonian? (A similar  problem was analysed by
Kijowski in
\cite{Kij} and also in \cite{Silva,Nester}). In our case
the answer
has to be looked for  into the way we consider the variation of the
Hamiltonian. Namely, we can consider different one-parameter families
of solutions each one containing the same solution (i.e. in the
infinite dimensional space of solutions we consider different curves
passing through the same point). Different one--parameter families of
solutions (i.e. different curves) correspond to different ways in
which the variation of the Hamiltonian is carried out and, in turn,
to different definitions  of energy.

The plan of the paper is as follows.
In Section \ref{Nother Theorem} we shall present the theoretical
framework leading to the definition of Hamiltonian and energy in
relativistic field theories.
 We stress that even though the formalism will  be presented
with the aim of being  applicated to  General Relativity, it will be
developed  inside a geometrical  framework which is common to all
natural as well as to all gauge--natural relativistic theories so
that, at least in principle, all classical theories of physical
interest  can be covered.  Starting from Section \ref{General
Relativity} we shall focus only on General Relativity
which for simplicity  will  be considered only  in vacuum. 
The generalization to cases with matter is an easy task which adds
nothing to the understanding of the problem and, as such,  we leave
it to some interested reader.
 It is in fact the second order Hilbert
Lagrangian (as well as any Lagrangian  of order
$k>1$) which  is sensitive to the formalism  presented here; first
order matter Lagrangians can in fact be treated in the usual way; see
\cite{Remarks,CADM}. First of all we shall shortly review
the geometry of the non--orthogonal foliation of (a portion of)
spacetime and we shall fix the notation. Then, in subsection
\ref{Hamiltonian}, the variation of the Hamiltonian together with 
Hamilton equations are calculated in order to test the formalism. The
quasilocal energy of the gravitational   field, defined as the
on--shell value of the Hamiltonian, is analysed in subsection
\ref{energy}. It will be there shown that   the variation of the
energy can be formally integrated provided  Dirichlet boundary
conditions are imposed. In this way a definition of quasilocal
\emph{internal} energy for a spatially bounded gravitating system is
obtained whence  the reference background subtraction terms are
properly taken into account. Moreover, a  comparison with the
alternative definitions  of internal energy  given in
\cite{Booth,HHnon,Kij} is carried out and a  (conditioned)
correspondence 
 is provided.
Finally, in subsection \ref{entropy}, the first principle of
black-hole thermodynamics is easily obtained as a trivial
consequence  of the cohomological properties of N\"other charges.

%%%%%%%%%%%%%%%%%%%%%%%%%%%%%%%%%%%%%%%%%%%%%%%%%%%%%%%%%%%%%%%
\section{N\"other Theorem}
\label{Nother Theorem}
We shall here consider a field theory described by a 
Lagrangian of order $k$. 
According to  the geometric approach  to  field theory (see for
instance
\cite{Lagrange,Trautman,Saunders,Sarda}) we are assuming that 
the configuration bundle of the  theory is
a bundle
$(Y,M,\pi)$. The base manifold
$M$, with $\dim M=m$,  is the space of parameters of the theory.
From now we shall identify the manifold $M$ with the
physical spacetime.  Fibered  adapted coordinates
on
$Y$ are
$(x^\mu,y^i)$, where
$\mu=1,\dots,  m=\dim M$, 
$i=1,\dots, n=\dim Y-\dim M$. A
\emph{configuration} is a (local) section $\s: M\ram Y$, locally
$x^\l\mapsto (x^\l,y^i=\s^i (x^\mu))$, which  to each
point
$x\in M$ associates the values of the fields ``$y$" (of components $
y^i $).  Thence, a configuration  completely describes the
evolution of the fields which represent a given physical model.
The fibered
manifold $(J^kY, \pi^k,M)$ denotes the $k$ jet prolongation
of the configuration bundle (i.e.  the bundle where fields live
together with their spacetime derivatives  up to order $k$
included) and
$j^k\s:M\ram J^kY$  denotes the jet prolongation of the section $\s$ 
(i.e. $j^k$ denotes derivation  up to order $k$).  Natural fibered
coordinates  on
$J^kY$ are obviously defined  and denoted by
$(x^\mu, y^i, 
\dots,y^i_{\mu_1\dots\mu_k})$.
In the framework so far developed a  Lagrangian of order $k$
can be seen as an horizontal form (or equivalently as a base
preserving fibered morphism) $ L: J^k Y \ram \Lambda^m(M)
$. 
It can be locally  written as
$L=\L (y^i, \dots,y^i_{\mu_1\dots\mu_k})\, ds 
$
where $\L$ is  a scalar density, called the \emph{Lagrangian density},
and
$ds=dx^1\wedge\dots\wedge dx^m$ is the local volume element  of
$m$--dimensional spacetime. Once the
Lagrangian is evaluated on (the jet prolongation of ) a
section it gives rise to a $m$--form $(j^k\s)^*
L$ on  spacetime $M$.

 A vertical vector field  $X$ is a
section of the vertical bundle
$VY\ram Y$, i.e. it is 
 a vector field on the configuration bundle $Y$ which is
everywhere tangent to the fibers (or, equivalently, projects down to
zero under  the tangent map of the bundle projection).
 Locally, it  reads as  $X=X^i \partial/\partial
y^i=(\delta y^i )\partial/\partial
y^i$ and it describes what is classically called  the ``variation''
$\delta y^i$ of the dynamical fields. The $k$--jet prolongation $J^k
X$ is a section of the bundle $J^k(VY)\ram Y$ and, roughly speaking,
it describes  the variation $\delta y^i, \delta y^i_\mu,
\dots \delta y^i_{\mu_1\dots \mu_k }$ of the dynamical fields
together with their derivatives up to order $k$.

Given a vertical vector field $X$ and a section $\s$ and denoting by
$\phi_\tau$ the (local) flow of $X$ we define a one--parameter family
$\s_\tau$ of sections through the rule $\s_\tau=\phi_\tau\circ\s$.
The variation $\delta_X \o$ with respect to $X$ of a differential form
$\o$ on any $J^hY$ ($h\ge 1$) is defined according to the rule 
\[
\delta_X \o=\left.{d\over d\tau} \o(j^h \s_\tau)\right \vert_{\tau=0}
\]

From now
on, for notational simplicity, we shall assume that the
Lagrangian, as well as all horizontal forms on any
prolongation of the configuration bundle, are already computed
along sections so they actually define differential forms on
spacetime, e.g. $L$ stands shortly for $(j^k\s)^* L$. In other
words we use the shorcut  $y^i$ to actually denote
$y^i=\s^i(x)$, i.e. the $i$--th component  of the section $\s$ with
respect to the $i$--th fibered coordinate $y^i$. Analogously
$y^i_\mu$ is used instead of $\partial_\mu \s^i(x)$ and so on for the
derivatives
$j^h y$ of higher order. Nevertheless we emphasize that all
calculations   we shall carry on  in the standard language
of differential calculus on the base manifold
$M$ can be alternatively carried on at the level of jet
bundle calculus and only at the end
the results so obtained can be pulled back on spacetime. In the
latter case globality and uniqueness problems are easily solved and
calculations can be algorithmically defined. Since  we
are here interested in applications, we shall skip the rigorous
geometric framework and  we refer the reader
to
\cite{Remarks,Robutti} for a deeper insight into the
mathematical details of the matter. To stimulate  the interest of
physically--oriented  readers we shall keep as much as possible the
formal setting  to a minimum and we shall use  a coordinate language.

Given a vertical vector field $X$, the variation $\delta_X \L$ of the
Lagrangian
$L=\L(j^k y)\,ds$ can be generally written, through a well known 
integration by parts procedure, as follows:
\begin{equation}\delta_X \L(j^{k} y)=e_i(j^{2k} y)X^i +d_\l
F^\l(j^{2k-1} y,j^{k-1}X)\label{FVFuno}
\end{equation} 
or, in terms of differential forms, as
\begin{equation}\delta_X L=e(L,X)+d F(L,X)\label{FVF}
\end{equation} 
Here the  $m$--form
\be
e(L,X)=e_i(j^{2k} y)X^i\,ds
\ee
and the $(m-1)$--form
\be
F(L,X)= F^\l(j^{2k-1}
y,j^{k-1}X)\,ds_\l\,, \qquad ds_\l=i_\l\, ds
\ee
are called, respectively, \emph{the Euler--Lagrange  form} and
\emph{the Poincar\'e--Cartan  form}. We stress  that the
Poincar\'e--Cartan  form $F(L,X)$ depends  linearly on 
$X$ together with  its derivatives up to the order $k-1$ included. [We
also remark that, for higher order theories    the
Poincar\'e--Cartan  form is not unique since it usually depends on
the choice of a  linear connection on spacetime; see
\cite{FPC}. We shall not care about this in the following since we
shall only deal  with General Relativity where a canonical
Poincar\'e--Cartan form can be chosen.]

 The \emph{critical
sections}
$\s:x^\l\mapsto y^i=\s^i(x)$ are those which  satisfy the
Euler--Lagrange equations: 
\begin{equation}\left.e_i(j^{2k}y)\right\vert_{y=\s(x)}=0
\end{equation} 

We shall now consider \emph{natural theories},
namely, theories describing  the behavior of geometric objects (such,
e.g.,  tensor fields, tensor densities or linear connections)
by means of  Lagrangians which are  covariant with respect to the
action  of diffeomorphisms of spacetime
$M$. In mathematical language natural theories are the ones
fulfilling    the
\emph{fundamental identity} 
\begin{equation}
d(i_\xi L)=\left\{ {\partial \L\over \partial y^i}
\pounds_{\xi} y^i+\dots+
 {\partial \L\over\partial  y^i_{\mu_1\dots\mu_k}}
\pounds_{\xi}y^i_{\mu_1\dots\mu_k}
\right\}ds
\label{simmetriainfinitesima} 
\end{equation} 
for each vector field $\xi$
on spacetime; see \cite{Lagrange}.
Inserting  the first variation formula \form{FVF} (with $X$ replaced
by $\pounds_\xi y$) into the right hand side of the fundamental
identity
\form{simmetriainfinitesima} we obtain 
the conservation laws
\begin{equation}
d\E(L,\xi)=\W(L,\xi)\label{WU}
\end{equation} 
where the ($m-1$)--form $\E(L,\xi)$  and the
$m$--form $\W (L,\xi)$ are  defined, respectively, as
follows:
\begin{eqnarray}
&&\E(L,\xi)=F(L,\pounds_\xi y) -i_\xi L\label{noethercurrent}\\
&&\W(L,\xi)=-e(L,\pounds_\xi y)
\end{eqnarray}
and are called, respectively, \emph{the N\"other current} of
$L$ relative to $\xi$ and \emph{the work current}. Notice
that, by definition, one has $\left.\W(L,\xi)\right\vert_{y=\s(x)}=0$
whenever  $\s$ is a solution of field equations, so that $\W(L,\xi)$
evaluates by \form{WU} the ``work'' performed  off--shell, i.e. by
sections which are not critical. Accordingly, from equation
\form{WU} we infer that  the differential form 
$\E(L,\xi)$ is conserved (i.e. closed) on--shell.  Moreover, in all
natural theories the  map $\xi\mapsto \E(L,\xi)$ is a linear partial
differential operator in the coefficients $\xi^\mu$. Hence, the 
N\"other current $\E(L,\xi)$, through a (covariant) integration by
parts,   can be rewritten as 
\begin{equation}
\E(L,\xi)=\tilde \E(L,\xi)+ d\U(L,\xi)\label{tildeepiudU}
\end{equation} 
where the $(m-1)$--form $\tilde \E(L,\xi)$  is
called \emph{the reduced current} and it is vanishing  on--shell
because  it is proportional  to a combination of field equations; see
\cite{Robutti,Remarks}. The $(m-2)$--form $\U(L,\xi)$  is
instead called
\emph{the superpotential}.  It depends on the fields $y$ and
their derivatives up to order $2 k-2$ and it is linear  in the
components $\xi^\mu$ and their derivatives up to order $k-2$.
Hence,  it follows from \form{tildeepiudU} that the N\"other current
$\E$ is not only closed  but it is also exact on--shell.\footnote{We
stress that the N\"other current 
 as well as the
superpotential are \emph{algorithmically} and \emph{uniquely}
defined at the bundle level in terms of jet bundle morphisms and they
are 
\emph{canonically} associated to the Lagrangian. 
\emph{Then} they are computed along a configuration  $\s$ and
they  give rise to differential forms on spacetime.}
Since $\E(L,X)$ is an ($m-1$)--form in spacetime, it
can be  integrated over an $(m-1)$--dimensional region $\Sigma$,
namely, a    submanifold $\Sigma$ of $M$ with a boundary  $\pa \Sigma
\subset 
\Sigma\subset M$ which, in   turn, is a 
compact  $(m-2)$--dimensional submanifold. With this formalism at
hands, one could be tempted  to define the N\"other charge
$Q_\Sigma(L,\xi, \s)$  along a section
$\s$ and relative to $\xi$  as follows:
\ba
Q_\Sigma(L,\xi, \s)&=&\int_\Sigma
\left.\E(L,\xi)\right\vert_{y=\s(x)}\nonumber\\
&=&\int_\Sigma
\left.\tilde \E(L,\xi)\right\vert_{y=\s(x)}+\int_{\partial \Sigma}
\left.\U(L,\xi)\right\vert_{y=\s(x)}
\label{Qnoncorretta}
\ea
 (notice that the
N\"other charge $Q_\Sigma(L,\xi, \s) $  becomes  a pure
boundary integral
 whenever $\s$ is a
solution of field equations). Nevertheless it is  well
known 
that the definition given in this way does not fit the physical
expectation values.

\begin{Example}
\label{esempio}{\rm
The Hilbert Lagrangian 
\be
L_H={1\over 2\kappa}\sqrt{g} \,g^{\mu\nu}\,R_{\mu\nu}\,ds
\ee
(with $\kappa= 8\pi$ in geometric units with  $c=G=1$)
is a second order Lagrangian ($k=2$) on the  configuration bundle
 $Lor(M)\ram M$ of Lorentzian metric on the $4$
dimensional spacetime $M$. A vertical vector field on the
configuration bundle reads locally as $X=(\delta g_{\mu\nu})\,
\partial/\partial g_{\mu\nu}$. The variation of the Lagrangian
can be written as:
\be
\delta L_H=-{1\over 2\kappa}\sqrt{g} \,G^{\mu\nu}\,\delta
g_{\mu\nu}\,ds+d\left[ {\sqrt{g}\over 2\kappa}g^{\mu\nu}\delta
u^\al_{\mu\nu}\, ds_\al
\right]
\ee
where 
\be
u^\al_{\mu\nu}=\Ga \al\mu\nu-\delta^\al_{(\mu}
\Gamma^\s_{\nu)\s}\,\qquad G_{\mu\nu}=R_{\mu\nu}-{1\over
2}g_{\mu\nu}\,R
\ee
Hence, the Eule--Lagrange form and the Poincar\'e--Cartan form are
given, respectively, by
\ba
&&e(L_H,X)=-{1\over 2\kappa}\sqrt{g} \,G^{\mu\nu}\,\delta
g_{\mu\nu}\,ds\\
&&F(L_H,X)={\sqrt{g}\over 2\kappa}g^{\mu\nu}\delta
u^\al_{\mu\nu}\, ds_\al
\ea
The N\"other current, defined as 
\[
\E(L_H,\xi)={1\over 2\kappa}\left\{
g^{\mu\nu}\pounds_\xi(
u^\al_{\mu\nu})-\xi^\al\,L_H\right\} ds_\al
\]
 splits as
$\E(L_H,\xi)=\tilde \E(L_H,\xi)+d\U(L_H,\xi)$ where 
\ba
&&\tilde \E(L_H,\xi)={\sqrt{g}\over \kappa}G^\al_\nu\,\xi^\nu ds_\al\\
&&\U(L_H,\xi)={\sqrt{g}\over 2\kappa}g^{\nu\r}\,\nabla_\r \xi^\mu\,
ds_{\mu\nu}
\ea
Notice that the superpotential $\U(L_H,\xi)$ is noting but the Komar
superpotential; see \cite{Komar}. 
When  integrated on spacelike hypersurfaces,
it is well--known that
the Komar superpotential does not reproduce the expected values for
all conserved quantities.
Indeed it is affected by the so--called ``anomalous factor problem''
when computed along asymptotically flat solutions (see \cite{Katz})
or, even worse, it gives rise to divergence problems on solutions of
Einstein's equations not featuring an asymptotically flat behavior
(see \cite{BTZ,TaubBolt}). 

This example strongly suggests that a modification 
in the definition of N\"other charges has to be provided.
\CVD}\end{Example}

Let us consider  again a  general natural
theory. 
According to  definition
\form{noethercurrent}, for each vertical vector field $X=(\delta
y^i)\,\partial/\partial y^i$ we have:
\ba
\delta_X \E(L,\xi)&=& \delta_X F(L,\pounds_\xi y)-i_\xi
\,(\delta_X L)\um^{\form{FVF}}
\delta_X F(L,\pounds_\xi y)-i_\xi e(L,X)-i_\xi
dF(L,X)\nonumber\\ &=& \delta_X F(L,\pounds_\xi y)-\pounds_\xi 
F(L,X)-i_\xi e(L,X)+d i_\xi F(L,X)\nonumber\\
&=&\o(L,X,\pounds_\xi y)-i_\xi e(L,X)+d\,( i_\xi
F(L,X))\label{deltaE}
\ea
where \emph{the symplectic current} $\o(L,X,Y)$ relative to $L$
and calculated for two vertical vector fields $X$ and $Y$ is
defined as (see \cite{Waldsymp}):
\be
\o(L,X,Y)=\delta_X F(L,Y)-\delta_Y  F(L,X)\label{symplectic}
\ee
Equation \form{deltaE} suggests to redefine the variation
$\delta_X \hat Q$ of
 the corrected conserved quantity $\hat Q$ as follows
\ba
\delta_X \hat Q_\Sigma(L,\xi)&=&\int_\Sigma\left[ \delta_X \E(L,\xi)
-d(i_\xi F(L,X))\right]\nonumber\\
&\um^{\form{tildeepiudU}}&\int_\Sigma
\delta_X
\tilde\E(L,\xi)+\int_{\partial \Sigma}\left[\delta_X \U(L,\xi)-
i_\xi F(L,X)\right]\label{deltatildeq}
\ea
Hence, from equation \form{deltaE} we obtain:
\be
\delta_X \hat Q_\Sigma(L,\xi)=\int_\Sigma\left\{\o(L,X,\pounds_\xi
y)-i_\xi e(L,X)\right\}\label{12bis}
\ee
This prescription is at the basis of the so--called ADM
\emph{covariant} formalism; see \cite{Remarks,CADM}, as well as it is
at the basis of a Hamiltonian analysis through a symplectic viewpoint
(see
\cite{Wald,Wald95,Waldsymp}). Indeed, whenever 
the (nowhere vanishing) vector field $\xi$
is  transverse to a Cauchy hypersurface $\Sigma$ we define  the
(variation of the) Hamiltonian $\hat H(L,\xi,\Sigma)$ simply as 
follows:
  \be
\delta_X \hat H(L,\xi,\Sigma)=\delta_X \hat
Q_\Sigma(L,\xi) \label{definitionH}
\ee
i. e. we identify the variation of the Hamiltonian  with the
variation  of the 
\emph{corrected} N\"other charge  $\hat Q_\Sigma$ relative to $\xi$.
First of all notice  that  definition \form{definitionH} has the
advantage of being independent on divergence terms added to the
Lagrangian. Indeed if we consider a new Lagrangian $L'$ differing
from $L$ by a total divergence, i.e. $L'=L+ d \b$, where  $\b$ is a
($m-1$)--form, we have 
\ba
&&\tilde \E(L',\xi)=\tilde
\E(L,\xi)\nonumber\\
&&\U(L',\xi)=\U(L,\xi)+ i_\xi \b\nonumber\\
&&i_\xi F(L',X)=i_\xi
F(L,X)+ i_\xi \delta_X\b\nonumber
\ea
 and hence $\delta_X \hat Q_\Sigma(L',\xi,)=\delta_X \hat
Q_\Sigma(L,\xi)$; see definition
\form{deltatildeq}. This means that divergence terms added to the
Lagrangian  just affect the superpotential  so that this fact would
reverberate on the definition of the Hamiltonian if it were
naively defined through the non corrected N\"other charge
\form{Qnoncorretta}. Namely, different divergence terms  would imply 
different definitions of Hamiltonian and, in turn, different
``boundary Hamilton equations''. Instead, the boundary term
$i_\xi F(L,X)$ added in the modified  definition \form{deltatildeq}
exactly counterbalances  the modifications in the superpotential and
we end up with a unique definition of Hamiltonian (unique inside
the set of Lagrangians differing from each other only by divergence
terms). This property is clearly welcome. The duty of the
Hamiltonian should in fact be  to dictate, via Hamilton equations, the
evolution laws of the dynamical fields in such a way that  Hamilton
equations have to correspond to (the  space+time decomposition of)
Euler--Lagrange equations. As well as these latter equations  are
insensitive to Lagrangian divergence terms one would expect the
same property to hold true also for Hamilton equations.

Notice also that, according to definitions
\form{deltatildeq} and \form{definitionH}, the (variation of the)
Hamiltonian consists of a  bulk term plus a  surface term. 
It is exactly  the surface term
which plays a key role
in  testing the viability  in the
 definition of the Hamiltonian. Indeed the variation of the
Hamiltonian is well--defined  if all surface terms in $\delta H$ 
vanish. That is, the surface terms in the Hamiltonian, when varied, 
should  exactly cancel out  surface terms arising in the
variation of the bulk term. In  this way, the variation of the
Hamiltonian  would be well--defined  on the full phase space 
of the dynamical fields rather than just on the restricted
space  of fields satisfying suitable boundary conditions
(namely, the boundary conditions necessary  to cancel out
possible  surface terms in
$\delta H$). 
Hence, in order to give a viable definition of Hamiltonian,
  we   propose here an additional boundary modification
to the general prescription
\form{deltatildeq}. We stress that, when dealing with  General
Relativity, the following modification will turn out to be relevant
only  when  the foliation of (a region of) spacetime does not feature
orthogonal intersections with  the timelike boundary. 

 Let us thence consider a (portion of a)  Cauchy hypersurface
$\Sigma$ in spacetime. Since the Poincar\'e--Cartan form
$F(L,X)$ is linear in
$X$ and its derivatives, owing to definition \form{symplectic} the
integral 
$\int_\Sigma \o(L,X,\pounds_\xi y)$ of the 
symplectic form on the hypersurface
$\Sigma$, by means of an integration by parts procedure,
can be rewritten  as follows:
\be
\int_\Sigma  \o(L,X,\pounds_\xi
y)=\int_\Sigma\tilde\o(L,X,\pounds_\xi y)+\int_{\partial \Sigma}
\tau(L,X,\pounds_\xi y)\label{24}
\ee
i.e. it splits into a bulk term plus a pure boundary term which,
 for $k>1$, is in general non vanishing
(and we shall see below that this  is
indeed the case for General Relativity). Accordingly, our  proposal is
to slightly modify definition
\form{deltatildeq} by pushing  the boundary term
$\int_{\partial \Sigma}
\tau$  directly into the definition of the Hamiltonian and defining
then  a new surface--adapted  Hamiltonian $H(L,\xi,\Sigma)$.
That is, we define the variation
$\delta_X H$ of the new Hamiltonian as follows
\ba
\delta_X H(L,\xi, \Sigma)&=&\int_\Sigma\left\{ \delta_X \E(L,\xi)
-d\left[i_\xi F(L,X)+\tau(L,X,\pounds_\xi
y)\right]\right\}\label{deltaH}\\ 
&=&\int_\Sigma
\delta_X
\tilde\E(L,\xi)+\int_{\partial \Sigma}\left[\delta_X \U(L,\xi)- i_\xi
F(L,X)-\tau(L,X,\pounds_\xi y)\right]\nonumber
\ea
This prescription is very close  in spirit to the original idea of
Regge--Teitelboim (see \cite{RT}) even though we are now dealing
with any natural field theory and the prescription is based on
a symplectic analysis of N\"other theorem instead of being based on 
a canonical analysis  of the ADM Hamiltonian. It    follows in fact
from
\form{12bis} and \form{24} that:
\be
\delta_X H(L,\xi, \Sigma)=\int_\Sigma \tilde\o(L,X,\pounds_\xi
y)-\int_\Sigma i_\xi e(L,X)\label{deltaxdiH}
\ee
and   equation \form{deltaxdiH}  ensures  that $\delta_X
H$  does not contain surface integrals which, as
already explained, would correspond  to Hamiltonian
boundary conditions on the dynamical fields. 

Notice also that the definition \form{deltaH} does
not alter the property of invariance of $\delta H$ with respect to the
addition of divergence terms into the Lagrangian.
Nevertheless the definition given depends in
general on the particular hypersurface $\Sigma$ we are dealing with.
Additional  boundary hypotheses have then  to be imposed in order to
avoid the problem. We refer the reader to \cite{Waldsymp} where the
dependence of the symplectic current on the Cauchy hypersurface  was
carefully investigated. We shall here skip the details on the matter 
and we shall instead focus on the definition of energy which is our
main concern in this paper.

We define the variation of the 
\emph{energy}
$E(L,\xi)$ relative to
$\xi$ as the on shell value of the variation of the Hamiltonian.
Hence,  the variation $\delta_X E(L,\xi)$ is obtained evaluating 
\form{deltaH} along solutions of field equations and considering 
variations with respect to a vertical vector field
$X$ which, in turn,  is a solution of linearized field equations,
which in our formalism are equivalent  to the equations 
\be
\delta_X\tilde
\E(L,\xi)=0
\ee
 Since under 
these hypoteses we have
 $i_\xi e(L,X)=0$, 
from definition  \form{deltaH} we obtain that $\delta_X
E(L,\xi)$ turns out to be a pure boundary term, i.e.
\be
\delta_X E(L,\xi)\doteq \int_{\partial \Sigma}\delta_X
\U(L,\xi)- i_\xi F(L,X)-\tau(L,X,\pounds_\xi
y)\label{deltaenergia}
\ee
while  equation \form{deltaxdiH} reduces to 
\be
\delta_X E(L,\xi)=\int_\Sigma\tilde\o(L,X,\pounds_\xi
y)\label{eqenergia}
\ee

Notice however that the expressions \form{deltaH} and
\form{deltaenergia} define only the variation of the
Hamiltonian and of the energy, respectively. The
problem arises whether  those expressions can be formally integrated
or not.
We shall see that, at least for General Relativity, this problem
admits a positive answer. Roughly speaking, the problem of
integrability  is related to the way in which  we keep the field
variations or, in other words, is related to the way we impose
boundary conditions (e.g.\ Dirichlet or Neumann conditions). Notice 
that in the present framework for boundary conditions we actually
mean  conditions which have to be satisfied by the vertical vector
fields
$X$ on the boundary
$\partial \Sigma$ (for example $\left.X\right\vert_{\partial
\Sigma}=0$). Different classes of vector fields correspond to
different sets of boundary conditions and they  will give rise to
different definitions of energy, as it is physically expected. Notice
that  boundary conditions are not imposed, a priori, by
adding divergences terms to the  Lagrangian we started from.
As already outlined in the Introduction, this would be a rather
unpleasant strategy mainly because we would have to modify the theory
from the very beginning, i.e. at the Lagrangian level,  accordingly on
what  kind of energy we are interested in (e.g. internal energy, free
energy and so on). Indeed we stress again that  each kind of energy is
deeply related  to its own boundary conditions (see
\cite{Kij,Nester,Silva}) and in the  Hamiltonian formulations based on
the canonical reduction of the action functional, boundary
conditions are dictated  by divergence terms added to the
Lagrangian and vice versa. 
Hence, in order  to satisfy different  boundary conditions we
would have to add  different divergence terms each time.
In other words,  each symplectic control--response boundary structure
dictates its own action functional.

 On the
contrary our N\"other--based prescription is insensitive on divergence
terms so that definition
\form{deltaenergia} is unchenged if we remain inside  any class
$[L]$ of Lagrangians, the elements of which consist of Lagrangians 
differing from each other only by divergence terms. Different
physical meanings can be nevertheless attributed to the  
formula
\form{deltaenergia}, which is the same for   each element of
the class
$[L]$, by considering variations with respect to different
classes of vertical vector fields $X$. 
Namely, boundary conditions  play an effective role  only at
the Hamiltonian level. The vector field $X$ (together with  its
jet prolongation) keeps track of  what has to  be held fixed on
the boundary, i.e.  it determines the control variables.
In this way  the Lagrangian
formulation  gets rid of a direct physical meaning which is
instead restored at the Hamiltonian level. Indeed, it is only
 in the Hamiltonian framework   where we read off  the
symplectic structure of the theory, namely the
control--response variables and the corresponding energetic
content; see \cite{Kij,Nester}. We can completely ignore  divergence
terms and the corresponding boundary conditions in the Lagrangian
formulation provided we assign  to the Lagrangian the only duty to
frame the theory in a well--defined geometric framework and to
provide the equations of motion\footnote{
Notice that, in the geometric approach to  Lagrangian theories, the
equations of motion  are mathematically recovered  from the variation
$\delta_X A_D(\s)$ of the action functional by postulating  that
critical sections  $\s$ are the ones  extremising  the action  for
\emph{any} region $D$  and for \emph{any} vector field $X$ with
compact support in $D$.  Hence all boundary  integrals  in $\delta_X
A_D(\s)$ are vanishing  due to the  hypothesis made on $X$. In this
way  action functionals  differing from each other  by boundary terms
give rise to the same equations of motion.
}. Instead, at the Hamiltonian level,  we
are free to  fix  the boundary conditions  in any way we want
through the choice of appropriate vector fields.

The application to General Relativity will help us to clarify
in detail how the idea works on.
%%%%%%%%%%%%%%%%%%%%%%%%%%%%%%%%%%%%%%%%%%%%%%%%%%%%%%%%%%%%%%%%%

\section{General Relativity}
\label{General Relativity}
As an application of the formalism let us now consider again 
standard General Relativity in vacuum with
$\dim M=4$ (notice however that the formalism goes through for
each spacetime dimension $m>2$. The case  $m=2$ will be instead
treated separately in a forthcoming paper \cite{forth}).

Let us consider a
$4$--dimensional region
$D\subseteq M$  which is diffeomorphic  to the
product $\Sigma\times \Re$  where $\Sigma$ is a 
$3$--dimensional closed  manifold  with boundary $
B=\partial \Sigma$. 
We denote this diffeomorphism by
\be
\psi:\Sigma\times \Re\ram D
\ee
For simplicity  we assume in the present Section that
the boundary
$B=\partial \Sigma$  is topologically a sphere (even though no
relevant change arises when we endow 
  $\Sigma$ with an internal boundary, i.e. when  the
manifold
$\Sigma$ is assumed to be the product
$S^2\times I$ where  $I$ is a closed real interval). For any 
$t\in \Re$  a hypersurface  $\Sigma_t\subset D$ is induced by $\psi$
according to the rule  $\Sigma_t=\psi (\Sigma\times
t)=\psi_t(\Sigma)$ and we require  it to be a spacelike Cauchy
hypersurface. Each map $\psi_t:\Sigma\ram \Sigma_t$ establishes
an embedding of
$\Sigma$ into $M$. The set of all $\Sigma_t$, for $t$ in
$\Re$,  defines a foliation  of $D$ labelled  by the time
parameter  $t$. Moreover, each $\Sigma_t$ intersects the
boundary $\partial D$ in a compact  $2$ dimensional surface 
$B_t$ which is diffeomorphic  to $B$, for all $t$ in $\Re$; see
Figure 1. The diffeomorphism is established  by the map
$\psi_t: B\ram B_t$. Hence the boundary  $\B=\partial D$  is a
timelike hypersurface  globally diffeomorphic to the product 
manifold $B\times \Re$. The time evolution  field $\xi$ in $D$
is defined through the (local) rule $\xi^\mu \nabla_\mu t=1$ and, on
the boundary  $\B$, it is tangent to the boundary itself. We
shall use Greek letters, ranging from 
$0$ to $3$,   to label indices of tensorial objects on $D$.
Tensors on $\Sigma$  are labelled  by Latin letters $a,b,\dots$
running from $1$ to $3$. Tensors  on $B\times \Re$ are labelled
by middle Latin letters $i,j,\dots=0,2,3$; while  Capital latin
letters $A,B,\dots=2,3$ are used for tensors on $B$. Notice
that tensorial objects $T^{\mu\nu\dots}_{\al\b\dots}$ on $M$,
with contravariant indices
$\mu,\nu\dots$  tangent to each $\Sigma_t$, can be pulled back 
on $\Sigma$  via the embedding maps $\psi_t$ and they originate
tensor fields $T^{cd\dots}_{ab\dots}$ on $\Sigma$. Accordingly
these tensor fields have a double interpretation. When labelled
by Greek indices  they are thought of as tensor fields  in 
spacetime $M$, whilst when they are labelled  by Latin indices  they
are instead thought of as  tensor on $\Sigma$. The same double
notation  applies, with the corresponding modifications,  for
tensor fields  which are ``tangent'' to $\B$  and tensor fields  on
$B\times \Re$; and also  for tensor fields ``tangent'' to  $B_t$ and
tensor fields on $B$ (where ``tangent''  is referred to each
controvariant index).
 
%%%%%%%%%%%%%%%%%%%%%%%%%%%%%%%%%%%%%%%%%%%%%%%%%%%%%%%%%%%%%
%\begin{figure}[t]
%\label{figura1}
%\vbox{
%\vskip 10.8truecm
%\special{illustration   Foliation.EPS [scaled 6000]}
%\vskip-98pt\hskip 4.8truecm
% \hskip 1truecm {\sl Fig.\ 1}
%\vskip-4.2truecm\hskip 4.2truecm $\bar n$
%\vskip-1.3truecm\hskip 4.4truecm $\bar u$
%\vskip 2.0truecm\hskip 9.2truecm $\B=\cup_{_t} B_t$
%\vskip-2.1truecm\hskip 9.0truecm $B_t$
%\vskip-1.6truecm\hskip 6.7truecm $\Sigma_t$
%\vskip-2.5truecm\hskip 5.5truecm $u$
%\vskip-.4truecm\hskip 4.2truecm $n$
%\vskip-1.2truecm\hskip 5.8truecm $D=\cup_{_t} \Sigma_t$
%\vskip6.5truecm
%\caption{non--orthogonal foliation of a spacetime  region
%$D$} }
%\end{figure}
%%%%%%%%%%%%%%%%%%%%%%%%%%%%%%%%%%%%%%%%%%%%%%%%%%%%%%%%%%%
We denote by $u^\mu$ the future directed  unit normal  to
$\Sigma_t$ and we denote by $n_\mu$  the outward pointing
unit normal of $B_t$ in $\Sigma_t$. Of course
$\left.u^\mu\,n_\mu\right\vert_{B_t}=0$. Notice that we are not
requiring that the hypersufaces $\Sigma_t$ intersect the
timelike boundary $\B$ orthogonally so that, in general, one has 
$\left.\xi^\mu\,n_\mu\right\vert_{B_t}\neq 0$. We also denote
by $\bar n^\mu$ the outward pointing
unit normal of $\B$ in $M$ and with $\bar u^\mu$ the future
directed  unit normal  of 
$B_t$ in $\B$. We have
$\left.\xi^\mu\,\bar n_\mu\right\vert_{B_t}=
\left.\bar u^\mu\,\bar n_\mu\right\vert_{B_t}=
0$. The evolution vector field can be decomposed as
$\xi^\mu=N\,u^\mu+N^\mu$ where the shift vector $N^\mu$ is tangent  to
the hypersurfaces $\Sigma_t$, i.e.
$ N^\mu\,u_\mu= 0$ (but in general 
$\left. N^\mu\,n_\mu\right\vert_{B_t}\neq 0$). The metrics
induced  on $\Sigma_t$, $\B$ and $B_t$  by the metric  
$g_{\mu\nu}$  are given, respectively, by:
\ba
&&h_{\mu\nu}=g_{\mu\nu}+u_\mu \,u_\nu\label{33}\\
&&\bar \g_{\mu\nu}=g_{\mu\nu}-\bar n_\mu \,\bar n_\nu\label{34}\\
&&\s_{\mu\nu}=h_{\mu\nu}-n_\mu \,n_\nu=\bar \g_{\mu\nu}+\bar
u_\mu
\,\bar u_\nu\label{35}
\ea
(Notice that $h_{ab}$, $\bar \g_{ij}$ and  $\s_{AB}$ will instead
denote the pull  back metrics on $\Sigma$, $B\times \Re$ and
$B$, respectively.) We shall denote by $h$, $\bar \g$ and $\s$,
respectively, the  absolute values  of the metric determinants.  
The metrics \form{33}, \form{34} and \form{35}  with an index raised 
through the contravariant metric $g^{\mu\nu}$ define the projection
operators  in the corresponding surfaces.
 We also denote by 
\ba
&&K_{\mu\nu}=-h^\al_\mu \nabla_\al u_\nu\\
&&\bar \Theta_{\mu\nu}=-\bar \g^\al_\mu \nabla_\al \bar
n_\nu\\
&&{\cal K}_{\mu\nu}=-\s^\al_\mu D_\al n_\nu
\ea
respectively, the extrinsic curvatures of  $\Sigma_t$ in $M$, 
of $\B$ in $M$ and  of $B_t$ in $\Sigma_t$.
The symbols $D$ and $\bar{\cal D}$ denote, respectively,   the
(metric) covariant derivative on $\Sigma_t$ compatible with
$h_{\mu\nu}$ and the covariant derivative on $\B$ compatible with the
metric
$\bar \g_{\mu\nu}$.
We shall denote  by $K=K_{\mu\nu} h^{\mu\nu}$,
$\bar \Theta=\bar \Theta_{\mu\nu} \bar \g^{\mu\nu}$ and 
${\cal K}={\cal K}_{\mu\nu} \s^{\mu\nu}$ the traces of the
appropriate extrinsic curvatures. The momentum
$P^{\mu\nu}$ of the hypersurface $\Sigma_t$ is defined as:
\be
P^{\mu\nu}={\sqrt{h}\over 2\kappa}\left( K
h^{\mu\nu}-K^{\mu\nu}\right)
\ee
while for the momentum $\bar\Pi^{\mu\nu}$ of the hypersurface
$\B$ we set:
\be
\bar\Pi^{\mu\nu}=-{\sqrt{\bar\g}\over 2\kappa}\left( \bar\Theta
\bar\g^{\mu\nu}-\bar\Theta^{\mu\nu}\right)
\ee
From a physical viewpoint   the surface
$B_t$ can be thought of as an istantaneous configuration of a
set of observers at the time $t$ and their history is
represented  by the timelike boundary $\B$; see \cite{Booth}. The
\emph{unboosted}, or unbarred,  observers are the ones  evolving  with
four velocity
$u^\mu$ and hence orthogonally  to the foliation $\Sigma_t$. They are 
at rest 
with respect to the
$\Sigma$ foliation.
They view $u^\mu$ and $n^\mu$ as  the unit normals  to $B_t$. 
We instead refer to the \emph{boosted}, or barred, observers to the
ones  which are  on the same surface $B_t$ at the istant of
time $t$  but  evolving with velocity $\xi^\mu$. 
They are comoving with
$\B$. Since
$\xi^\mu\in T\B$ then  boosted observers  view  $\bar u^\mu$ and
$\bar n^\mu$ as  the unit normal to $B_t$.
Barred and unbarred observers are related
by the boost relations:
\ba
&&\bar u_\mu=\g\, u_\mu +\g v \,n_\mu\\
&&\bar n_\mu=\g\, n_\mu +\g v\,u_\mu
\ea
where 
\be
\g v=\bar u_\mu n^\mu=- u_\mu \bar n^\mu = \sinh{\theta}
\ee
measures the non--orthogonality of the intersections of the
leaves $\Sigma_t$ with the timelike boundary $\B$. The
parameter $\theta $ is usually called the \emph{velocity parameter}.
Moreover we set 
$\g=(1-v^2)^{-1/2}$ where  
\be
v=-{\xi^\mu
n_\mu\over \xi^\mu u_\mu}= {N^\mu n_\mu \over N}\label{vv}
\ee
 is the boost velocity
between the observers. It measures the velocity in the
direction of
$n^\mu$ of an object with four velocity  $\xi^\mu$ as measured 
by the observer with velocity $u^\mu$, namely the
velocity in the $n^\mu$ direction of boosted observers with
respect to unboosted observers.

The vector field $\xi^\mu$, as already said,  can be
decomposed as $\xi^\mu=N\,u^\mu+N^\mu$. On the boundary $\B$ 
it can be also decomposed  as $\xi^\mu=\bar N\,\bar
u^\mu+\bar N^\mu$ where $ \bar N=N/\g$ and  $\bar
N^\mu=\s^\mu_\nu N^\nu$. From now on we shall use the notation
that a bar over a quantity will denote  that it is  defined
with respect  to  the normals $\bar u^\mu$ and $\bar n^\mu$, so
that, e.g., $\bar {\cal K}_{\mu\nu}$ would correspond to  the
extrinsic curvature of
$B_t$ embedded in a spacelike hypersurface with unit  normal
$\bar u^\mu$ in $B_t$ (or, roughly speaking, it is the extrinsic
curvature  of $B_t$ with respect  to the normal $\bar n^\mu$).
Moreover, the following identities hold true:
\be
 \sqrt{g}=N\,\sqrt{h}\, ,
\quad  \quad \sqrt{\bar\g}=\bar N \sqrt{\s}
\ee

\subsection{Hamiltonian}
\label{Hamiltonian}
Let us now consider the Hilbert Lagrangian $L_H$. 
According to the general prescription \form{deltaH} the
variation of the Hamiltonian is defined as follows:
\ba
\delta_X H(L_H,\xi, \Sigma_t)&=&
\int_{\Sigma_t}
\delta_X
\tilde\E(L_H,\xi)\label{newdef}\\
&+&\int_{B_t}\left[\delta_X \U_K(L_H,\xi)-
i_\xi F(L_H,X)-\tau(L_H,X,\pounds_\xi
g)\right]\nonumber
\ea
where
(see example \ref{esempio}) we set:
\ba
&&F(L_H,X)={\sqrt{g}\over 2\kappa}g^{\mu\nu}\delta
u^\al_{\mu\nu}\, ds_\al\label{43}\\
&&\tilde \E(L_H,\xi)={\sqrt{g}\over \kappa}G^\al_\nu\,\xi^\nu
ds_\al=\left\{N \H +N^\al\H_\al\right\} d^3\,x\label{constr}\\
 &&\U_K(L_H,\xi)={\sqrt{g}\over
2\kappa}g^{\nu\r}\,\nabla_\r
\xi^\mu\, ds_{\mu\nu}
\ea
and  $\H$ and $\H_\al$ in \form{constr} are called, respectively,   
the Hamiltonian constraint and the diffeomorphism constraints.

Now, our next task will  be  to calculate the term  $\tau(L_H,X,
\pounds_\xi\, g)$ and to rewrite
\form{newdef} in terms of quantities adapted to the spacetime
foliation. In this way it will be possible to establish whether 
formula
\form{newdef} is a viable definition for the Hamiltonian. Moreover  a
correspondence with other definitions existing in literature will be
possible.  From now on  we shall adopt  the notational choices made 
in
\cite{BLY} and we shall make repeated  use  of useful formulae there
demonstrated. Let us start with the relations
\ba
&&g^{\al\b}\delta u^\mu_{\al\b}\, u_\mu=2 \delta K+K^{\mu\nu}\delta
h_{\mu\nu}+D_\mu (h^\mu_\al \delta u^\al)\label{BLY1}\\
&&g^{\al\b}\delta u^\mu_{\al\b}\, \bar n_\mu
=2 \delta \bar \Theta+\bar \Theta^{\mu\nu}\delta
\bar \g_{\mu\nu}+\bar {\cal D}_\mu (\bar \g^\mu_\al
\delta \bar n^\al)\label{BLY2}
\ea
(the proof of which can be found in the aforementioned paper
\cite{BLY}). From
\form{43}, \form{BLY1} and Stokes' theorem it then follows that
\ba
&&\int_{\Sigma_t} F^\mu(L_H,X) ds_\mu={1\over
2\kappa}\int_{\Sigma_t}\sqrt{g}g^{\al\b}\delta u^\mu_{\al\b}ds_\mu=-
{1\over
2\kappa}\int_{\Sigma_t}g^{\al\b}\delta u^\mu_{\al\b}\,
u_\mu\, \sqrt{h}\, d^3x\nonumber\\
&&\phantom{int_{\Sigma_t}}=-
{1\over
2\kappa}\int_{\Sigma_t}\left\{2 \,\delta K+K^{\mu\nu}\delta
h_{\mu\nu}\right\}\, \sqrt{h} \, d^3x-
{1\over
2\kappa}\int_{B_t}\!n_\al h^\al_\mu \delta  u^\mu\sqrt{\s}
d^2x\nonumber\\ 
&&\phantom{int_{\Sigma_t}}=-
\int_{\Sigma_t}h_{\mu\nu} \delta P^{\mu\nu} d^3x -
{1\over
2\kappa}\int_{B_t}n_\mu \delta  u^\mu\sqrt{\s} d^2x\label{deltafH}
\ea
(notice that $\int_{\Sigma_t}\!\sqrt{g}\,f^\mu
\,ds_\mu=-\int_{\Sigma_t}\! f^\mu \, u_\mu\,\sqrt{h}\, d^3x$ for any
$3$--form $f=f^\mu\, ds_\mu$ and $\int_{\Sigma_t}D_\mu
X^\mu\, \sqrt{h}\, d^3x=\int_{B_t} X^\mu\, n_\mu \sqrt{\s}\, d^2x$ for
a vector field $X^\mu$). Let us now consider the integral of the
symplectic current
\form{symplectic} on the generic leaf  $\Sigma_t$ of the foliation.
From \form{deltafH} we obtain:
\be
\int_{\Sigma_t}\o (L_H,X, \pounds_\xi g)=
\int_{\Sigma_t}\tilde \o (L_H,X, \pounds_\xi g)+\int_{B_t}\tau
(L_H,X, \pounds_\xi g)
\ee
where
\ba
&&\tilde \o (L_H,X, \pounds_\xi g)=
\left\{(\pounds_\xi h_{\mu\nu})\,\delta P^{\mu\nu}
-(\pounds_\xi P^{\mu\nu})\,\delta
h_{\mu\nu}\right\}\,d^3\,x\label{tildeomega}\\
&&\tau(L_H,X, \pounds_\xi
g)={1\over 2\kappa}\left\{\pounds_\xi(\sqrt{\s}\,n_\mu\delta
u^\mu)-\delta(\sqrt{\s}\,n_\mu\pounds_\xi
u^\mu)\right\}d^2\,x\label{taudiL_H}
\ea
Notice that, having not imposed  the hypothesis of orthogonal
intersections, the $2$--form $\tau(L_H,X,\pounds_\xi g)$ does not
vanish at the boundary $B_t$ 
since $n_\mu$ in general is not surface--forming 
(instead, in presence of orthogonal boundaries,  $n_\mu$  is
 surface--forming and hence $\delta n_\mu$ is proportional to $n_\mu$
so that $ n_\mu\,\delta u^\mu=-u^\mu\,\delta n_\mu=0$). Hence formula
\form{newdef} actually constitutes a generalization  of previous
definitions found in  literature and based on N\"other theorem; see
\cite{Wald95,Remarks,Waldsymp,Sinicco}.

The direct calculation of the expression \form{newdef} turns
out to be rather cumbersome. 
Fortunately many  results can be found elsewhere and we shall
only reproduce the final results. First of all the bulk term in
\form{newdef} has been already computed in \cite{Booth,BLY} even
though  it was originally calculated with respect to unbarred
quantities. In terms of barred quantities it reads as follows:
\ba
\int_{\Sigma_t}\delta_X \tilde \E(L_H,\xi,
\Sigma_t)&=&\int_{\Sigma_t}\left\{\delta N
\H +\delta N^\al\H_\al +[h_{\al\b}] \delta P^{\al\b}-[P^{\al\b}]
\delta h_{\al\b}\right\}d^3x\nonumber\\
&&-\int_{B_t}d^2x \left\{
\bar N\delta (\sqrt{\s}\bar \epsilon)-\bar
N^\al\delta(\sqrt{\s}\, \bar j_\al)
+{\bar
N\sqrt{\s}\over2}\bar s^{\al\b}
\delta \s_{\al\b}\right.\nonumber\\
&&\phantom{\int_{B_t}d^2x N }
\left.+{1\over \kappa}\left[\pounds_\xi(\sqrt{\s})\,
\delta (\theta) -\delta
(\sqrt{\s})\,  (\pounds_\xi\theta)\right]\right\}\label{dHb}
\ea
where 
\ba
&&\bar \epsilon={1\over \kappa}\bar {\cal K}\\
&&\bar j_\al=-{2\over \sqrt{\bar \g}}\, \s_{\al\mu} \bar
\Pi^{\mu\nu}\,
\bar u_\nu\\  &&\bar s^{\al\b}={1\over \kappa}\left[(\bar n^\mu \bar
a_\mu)
\s^{\al\b}-\bar {\cal K}\,\s^{\al\b}+\bar {\cal K}^{\al\b}\right]\quad
(\bar a_\mu=\bar u^\nu\nabla_\nu \bar u_\mu)
\ea
are, respectively,  the \emph{surface  quasilocal energy}, the
\emph{surface angular momentum}  and the \emph{surface stress
tensor},   as viewed by the boosted observers. They are all tensors on
the  boundary surface $B_t$ and, after integration on the surface,  
they  can be used to describe the stress energy--momentum content of
the gravitational field inside $B_t$; see \cite{BY,BYdue,Booth,BLY}
 and reference quoted therein. Instead
$[h_{\al\b}]$ and $\left[P^{\al\b}\right]$ in \form{dHb} are a
shortcut  for
\ba
 [h_{\al\b}]&=&{2\kappa\,N\over\sqrt{h}}(2 P_{\al\b}-
h_{\al\b} P)+D_\al N_\b+D_\b N_\al\label{equazADM1}\\
\left[P^{\al\b}\right]&=&-{N\sqrt{h}\over 2\kappa} ({\cal
R}^{\al\b}-{1\over 2} h^{\al\b} {\calÊR})+{\kappa\, N\over
\sqrt{h}}h^{\al\b}(P^{\mu\nu}P_{\mu\nu}-1/2P^2)\nonumber\\
&&-{2N\kappa\over\sqrt{h}}(P^{\al\mu}P^\b_\mu-1/2
P\,P^{\al\b})+\nonumber\\ 
&&+{\sqrt{h}\over 2\kappa}(D^\al D^\b
N-h^{\al\b}h^{\mu\nu}D_\mu D_\nu N)+\nonumber\\
&&+D_\mu(P^{\al\b} N^\mu)-D_\mu N^\al P^{\b\mu}-D_\mu N^\b
P^{\al\mu}\label{equazADM2}
\ea
having denoted by ${\cal R}_{\al\b}$ the Ricci tensor of the
$3$--metric $h_{\al\b}$.

Let us now consider the surface terms into
\form{newdef} to which we shall refer as $\delta_X
H_{_{\hbox{\scriptsize surf}}}(L_H,\xi,
\Sigma_t)$. We shall carry out  the calculation referring to the set
of barred observers (the ones comoving with
$\B$) instead of referring  to unbarred observers (at rest 
with respect to the
$\Sigma$ foliation).
We shall also make repeated use  of the property:
\be
\int_{B_t} \U=\int_{B_t}\!\! \sqrt{\s}\,d^2x\,
\U^{[\al\b]}n_\al u_\b=\int_{B_t}\!\! \sqrt{\s}\,d^2x\,
\U^{[\al\b]}\bar n_\al \bar u_\b
\ee
which holds true  for any $2$--form $\U=1/2\, \sqrt{g}\,
\U^{\al\b}\,ds_{\al\b}$.

 The first term in the surface integral 
\form{newdef}, after a bit of algebra (see
\cite{Wald95,BYOur}) can be written as follows 
\ba
\delta\left[ \int_{B_t}\U_K(L_H,\xi)\right]&=&{1\over
2\kappa}\int_{B_t}d^2 x\,\delta\left[ 2\sqrt \s \bar u^\mu
\bar \Theta ^\al_\mu
\xi_\al\right]\nonumber\\
&&+{1\over
2\kappa}
\int_{B_t}d^2 x\delta\left[\sqrt \s\bar u_\al \pounds_\xi \bar
n^\al\right]\label{PEZZO1}
\ea
For the second term in  $\delta_X H_{_{\hbox{\scriptsize
surf}}}(L_H,\xi,
\Sigma_t)$ we obtain instead:
\ba
-\int_{B_t} i_\xi F(L_H,\xi)&=&{1\over
2\kappa}\int_{B_t}\sqrt{\s} d^2 x \bar N\left[ 2\delta \bar
\Theta +\bar \Theta^{\mu\nu} \delta \bar
\g_{\mu\nu}\right]\nonumber\\
&& -{1\over
2\kappa}\int_{B_t} d^2 x \pounds_\xi\left[\sqrt{\s} \bar u_\mu
\delta \bar n^\mu\right]\label{PEZZO2}
\ea
The third surface term $\int_{B_t}\!\!\tau$ has already been written
down in
\form{taudiL_H}, i.e.
\be
-\int_{B_t}\tau
(L_H,X, \pounds_\xi
g)={1\over
2\kappa}\int_{B_t}\left\{-\pounds_\xi(\sqrt{\s}\,n_\mu\delta
u^\mu)+\delta(\sqrt{\s}\,n_\mu\pounds_\xi
u^\mu)\right\}d^2x\label{PEZZO3}
\ee
The value of  $\delta_X H_{_{\hbox{\scriptsize
 surf}}}(L_H,\xi,
\Sigma_t)$ results by the sum
of \form{PEZZO1}, \form{PEZZO2} and \form{PEZZO3}. In order to
gain some simplification we shall make use of the relations
(see \cite{BLY}): 
\ba
&&\bar\Theta_{\mu\nu}=\bar{\cal K}_{\mu\nu}+ (\bar n^\al 
\bar a_\al) \bar u_\mu \bar u_\nu+2\s^\al_{(\mu}\bar u_{\nu)}
\bar K_{\al\b} \bar n^\b\\
&&\delta \bar \g_{\mu\nu}=-(2/\bar N) \bar u_\mu\bar u_\nu\delta
\bar N -(2/\bar N)\s_{\al(\mu} \bar u_{\nu)}\delta \bar N^\al +
\s^\al_{(\mu}\s^\b_{\nu)}\delta \s_{\al\b}\\
&&n_\mu \delta u^\mu+\bar u_\mu \delta \bar n^\mu=-2\delta
\theta
\ea 
where
\be
\s^\al_{\mu}
\bar K_{\al\b} \bar n^\b=\s^\al_{\mu}
 (n^\b \, K_{\al\b}-\nabla_\al \theta)
\ee
After a little  of algebra we reach the final result:
\ba
 \delta_X H_{_{\hbox{\scriptsize surf}}}(L_H,\xi,
\Sigma_t)&=&\int_{B_t}d^2x \left\{
\bar N\delta (\sqrt{\s}\,\bar \epsilon)-\bar
N^\al\delta(\sqrt{\s}\,\bar j_\al)
+{\bar
N\sqrt{\s}\over2}\bar s^{\al\b}
\delta \s_{\al\b}\right.\nonumber\\
&&\phantom{\int_{B_t}d^2x N }
\left.+{1\over \kappa}\left[\pounds_\xi(\sqrt{\s})\,
\delta (\theta) -\delta
(\sqrt{\s})\,  (\pounds_\xi\theta)\right]\right\}\label{dHs}
\ea
Finally, the sum of the Hamiltonian bulk term \form{dHb} and the
Hamiltonian surface term \form{dHs} gives the result 
\be
\delta_X H(L_H,\xi, \Sigma_t)=\int_{\Sigma_t}\left\{\delta N
\H +\delta N^\al\H_\al +[h_{\al\b}] \delta P^{\al\b}-[P^{\al\b}]
\delta h_{\al\b}\right\}\,d^3x\label{explicitdeH}
\ee
We 
stress  that the boundary terms  arising  in the variation of the
bulk term  in
\form{dHb} exactly cancel out  the variation of the boundary term
\form{dHs}.

\vspace{1truecm}

In order to explicitly write down the Hamilton equations
\form{deltaxdiH} we still have to calculate  the term
\be
\int_{\Sigma_t} i_{\xi} \, e(L_H,X)=-{1\over
2\kappa}\int_{\Sigma_t}\sqrt{g} \, G^{\mu\nu} \delta
g_{\mu\nu}\,
\xi^\r\, ds_\r
\ee
Taking into account the relations 
\ba
&&\delta g_{\mu\nu}=-{2u_\mu u_\nu\over N} \delta
N-{2h_{\al(\mu}u_{\nu)}\over N}\delta
N^\al+h^\al_{(\mu}h^\b_{\nu)}\delta h_{\al\b}\nonumber\\
&&\int_{\Sigma_t}\sqrt{g}\, f\,\xi^\r\, ds_\r
=\int_{\Sigma_t}\sqrt{h}\,f \,N\, d^3x\nonumber
\ea
we have
\ba
&&
\int_{\Sigma_t}  i_\xi\, e(L_H,X)=\label{EulerH}\\
&&\phantom{dfdd}={1\over
2\kappa}\int_{\Sigma_t}\!\!\sqrt{h}\, d^3x\left[
2 G^{\mu\nu} u_\mu u_\nu\delta N+
2 G^{\mu\nu} h_{\al\mu} u_\nu\delta N^\al
-G^{\mu\nu}h^\al_\mu h^\b_\nu \delta h_{\al\b}
\right]\nonumber
\ea
Taking into account \form{tildeomega}, \form{explicitdeH} and 
\form{EulerH},
the Hamilton equations \form{deltaxdiH}, i.e.:
\be
\delta_X H(L_H,\xi, \Sigma_t)=\int_{\Sigma_t}
\tilde\o(L_H,X,\pounds_\xi y)-\int_{\Sigma_t} i_\xi
e(L_H,X)\label{analogy}
\ee
read as follows:
\ba
&&-{\sqrt{h}\over \kappa}G^{\mu\nu}u_\mu u_\nu=\H=0\label{HE1}\\
&&-{\sqrt{h}\over \kappa}G^{\mu\nu}h_{\al
\mu}u_\nu=\H_\al=0\label{HE2}\\
 &&\pounds_\xi
h_{\mu\nu}=[h_{\mu\nu}]\label{HE3}\\
 &&{\sqrt{g}\over
2\kappa}G^{\mu\nu}h^\al_\mu h^\b_\nu=\pounds_\xi
P^{\mu\nu}-[P^{\mu\nu}]=0\label{HE4}
\ea
Notice that equations \form{HE1}, \form{HE2} and \form{HE4} 
correspond to Einstein's equation  in the
$(3+1)$ formalism while \form{HE3} is nothing but the definition of
the momentum $P^{\mu\nu}$, i.e. the Legendre transformation.

Hence the prescription  \form{newdef}  originates  the correct
evolution equations  in the phase space $(h_{ab},P^{ab})$ of General
Relativity and it does not give rise to additional boundary
equations; see \cite{Kij,Sinicco}. \footnote{Notice, however, that 
Hamilton boundary equations would indeed arise
if the correction term $\int_{B_t}\!\! \tau$ in formula
\form{newdef} were missing.} For this reason the definition 
\form{newdef} can be truly considered as a genuine  generalization 
of the
Regge--Teitelboim Hamiltonian formulation
\cite{RT} to
arbitrary spacetime dimensions and to solutions with arbitrary
asymptotic behavior (and not necessarily stationary).

Notice also the close formal analogy of expression
\form{analogy} with the definition of Hamiltonian in Classical 
Mechanics. In Classical Mechanics one defines the generalized
Hamiltonian through the usual rule
$\left.H(q,p)\right\vert_{p(q,\dot q)}=p(q,\dot q)\,\dot q
-L(q,\dot q)$ so that 

\be
\delta H=\dot q\delta p-\dot p \delta q +\left(\dot
p-\partial L/\partial q\right)\delta q\label{hammeccanica}
\ee
In trying to establish
an  analogy between \form{hammeccanica} and definition 
\form{analogy} the term
$\int_{\Sigma_t}
\tilde \o(L_H,\xi, \Sigma_t)$ in \form{analogy} can be
seen as the 
$\dot q\delta p-\dot p \delta q$ term  -- see \form{tildeomega} --
while the term
$\int_{\Sigma_t} i_\xi e(L_H,X)$ clearly corresponds to
Lagrange equations $\left(\dot p-\partial L/\partial
q\right)\delta q$ -- see \form{EulerH}.

The question arises whether a Hamiltonian $H$  exists such that its
variation $\delta H$  equals the expression \form{explicitdeH}. The
answer to the problem is given by:
\ba
&&H(L_H,\xi, \Sigma_t)-H_0(L_H,\xi,
\Sigma_t)=\int_{\Sigma_t}\left\{N\H +N^\al
\H_\al\right\}d^3x
\label{Hexp}\\
&&\phantom{H(L_H,\xi, \Sigma_t)-H_0}+\int_{B_t}
\sqrt{\s}\left\{\bar N \bar {\epsilon} -\bar N^\al \bar
j_\al\right\}d^2x +{1\over \kappa}\int_{B_t}\pounds_\xi(\sqrt{\s})
\,\theta \,d^2x
\nonumber
\ea
where $H_0(L_H,\xi, \Sigma_t)$ turns out to be a constant of
integration in the left hand side of \form{explicitdeH}. [It will play
a fundamental role when dealing with the definition of energy since 
it will play the role of a reference term which defines the zero level
for the energy itself]. Indeed, the variation of
\form{Hexp} turns out to be:
\ba
\delta_X H(L_H,\xi, \Sigma_t)&=&\int_{\Sigma_t}\left\{\delta N \H
+\delta N^\al\H_\al +[h_{\al\b}] \delta P^{\al\b}-[P^{\al\b}]
\delta h_{\al\b}\right\}d^3x\nonumber\\
&&-\int_{B_t}\left\{\bar \Pi^{\mu\nu}\delta
\bar \g_{\mu\nu}-{1\over \kappa}\pounds_\xi(\theta\,
\delta
\sqrt{\s})\right\}d^2x\label{deltaXH}
\ea
and it gives  rise to expression \form{explicitdeH} provided  that,
during the variation,  the boundary metric $\bar \g_{\mu\nu}$  is kept
fixed. [We also remark  that it is indeed the Dirichlet condition
$\left.
\delta
\bar
\g\right\vert_\B=0$ together with the requirement that $\xi$ be a
boundary Killing vector field that ensures  the definition of $\delta
H$ to be independent on the choice of a particular Cauchy
hypersurface].

Moreover, expression \form{deltaXH}  can be used  to calculate
the time rate of change  of the Hamiltonian. To this end we have to
replace  the variations $\delta$ of the dynamical variables in
\form{deltaXH} with their Lie dragging $\pounds_\xi$ along the
evolution vector field
$\xi$. Assuming  that Einstein's equations are satisfied we obtain:
\be
\pounds_\xi\, H(L_H,\xi, \Sigma_t)=-\int_{B_t}\left\{
\bar \Pi^{\mu\nu}\pounds_\xi\,
\bar \g_{\mu\nu}-{1\over \kappa}\pounds_\xi(\theta\,
\pounds_\xi\,
\sqrt{\s})\right\} d^2x
\ee
which is vanishing if $\pounds_\xi\,\bar \g_{\mu\nu}=0$, i.e. if the
vector field $\xi$  is a boundary Killing vector field for the metric
$\bar \g$ or, in other words, if  the gravitational system is
isolated from the outside; see \cite{BY,BYdue,Booth}.
%%%%%%%%%%%%%%%%%%%%%%%%%%%%%%%%%%%%%%%%%%%%%%%%%%%%%%%%%%%
\subsection{Energy in General Relativity}
\label{energy}
Let us now consider   the definition  of  energy. It could be
obtained simply by evaluating \form{Hexp} on--shell. In order to
better understand  the role played  by the reference term $H_0$  and
its relations with boundary conditions let us  consider again
the on--shell  variation of the Hamiltonian. 
 We specialize definition \form{deltaenergia} for the Hilbert
Lagrangian $L_H$ so that $\delta_X E(g) =\delta_X
H_{_{\hbox{\scriptsize surf}}}(L_H,\xi,
\Sigma_t)$ and from \form{dHs} one has:
\ba
\delta_X E(g) &=&\int_{B_t}d^2x \left\{
\bar N\delta (\sqrt{\s}\bar \epsilon)-\bar
N^\al\delta(\sqrt{\s}\bar j_\al)
+{\bar
N\sqrt{\s}\over2}\bar s^{\al\b}
\delta \s_{\al\b}\right.\nonumber\\
&&\phantom{\int_{B_t}d^2x N }
\left.+{1\over \kappa}\left[\pounds_\xi(\sqrt{\s})\,
\delta (\theta) -\delta
(\sqrt{\s})\,  (\pounds_\xi\theta)\right]\right\}\label{Energy}
\ea
Notice that $\delta_X E(g)$ is a pure surface integral on
$B_t$. Nevertheless we stress that definition \form{Energy}
\emph{is not} a  quasilocal energy relative to the surface
$B_t$. Indeed, let us now consider a hypersurface
$\Sigma_t $ with an external boundary $B_t$ and an
internal boundary $C_t$, i.e. $\partial \Sigma_t=B_t
\cup C_t$. Taking  expression
\form{tildeomega} into
account,  equation \form{eqenergia} now  reads as:
\ba
\left.\delta_X
E(g)\right\vert_{\partial \Sigma_t}&=&\left.\delta_X
E(g)\right\vert_{B_t}-\left.\delta_X
E(g)\right\vert_{C_t}\nonumber\\
&=&\int_{\Sigma_t}\left[(\pounds_\xi
h_{\mu\nu})\,\delta P^{\mu\nu}-(\pounds_\xi P^{\mu\nu})\,\delta
h_{\mu\nu}\right]d^3x \label{kij}
\ea
For stationary solutions the right hand side vanishes together
with the correction term $\tau(L_H,X,\xi)$, so that
$\delta_X E$ has  the same numerical value when computed on different
surfaces belonging to the same homology class. In particular,
if spatial infinity is homologic to the surface $B_t$,
expression
\form{Energy} corresponds to the total energy or, better, it
corresponds to the total mass. 
Applications  to a number of specific solutions, including e.g.
Kerr--Newman, BTZ, Taub--bolt solutions, can be found in
\cite{Remarks,BTZ,TaubBolt}. In all the examples so far analysed  it
has been shown that  expression \form{Energy} (with $\tau=0$) always 
gives rise to the expected value, namely, the total mass of the
solution.

As previously remarked an
expression of the quasilocal energy can be instead  obtained
only when boundary conditions are imposed on the surface $B_t$.
 We define the \emph{quasilocal internal energy} $E(g,B_t)$ contained
in the surface $B_t$  to be the value obtained from
\form{Energy} by imposing Dirichlet metric boundary conditions.
Namely, 
 we consider a one parameter family of solutions of field
equations which admit the same  boundary metric
$\bar
\g_{\mu\nu}$ and we consider a vector field $X$ tangent to this
family, i.e. 
\be
\left.\delta \bar \g_{\mu\nu}\right\vert_{\cal B}=0\label{dirg}
\ee
Since $\left.\delta \bar \g_{\mu\nu}\right\vert_{\cal B}=0$ implies
$\left.\delta \bar N\right\vert_{\cal B}=0$, $\left.\delta
\bar N^\al\right\vert_{\cal B}=0$ and $\left.\delta 
\s_{\mu\nu}\right\vert_{\cal B}=0$, formula \form{Energy} can be
explicitly integrated:
\be
E(g,B_t)-E_0(g_0,B_t)=\int_{B_t}d^2x \left\{\sqrt{\s}\left(
\bar N\,\bar \epsilon-\bar
N^\al\, \bar j_\al\right)
+{1\over \kappa}\pounds_\xi(\sqrt{\s})\,
 \theta\right\}\label{e-e0}
\ee
We stress that  we are integrating along a family of solutions
satisfying the boundary condition \form{dirg} and thence the term
$E_0(g_0,B_t)$ corresponds to the quasilocal energy relative to a
solution $g_0$ inside the family itself, which then becomes the
reference point (or zero level) for the energy. Owing to the
properties 
$\left. \bar N=\bar N_0\right\vert_{\cal B}$, $\left.
\bar N^\al=\bar N^\al_0\right\vert_{\cal B}$ and $\left.
\s_{\mu\nu}=\s_{0\mu\nu}\right\vert_{\cal B}$, expression \form{e-e0}
can be equivalently rewritten as:
\be
E(g,B_t)=\!\!\int_{B_t}\!\!\! d^2x \!\!\left\{\!\sqrt{\s}\left[
\bar N(\bar \epsilon-\bar \epsilon_0)-\bar
N^\al(\bar j_\al-\bar j_{0\al})\right]
+{1\over \kappa}\pounds_\xi(\sqrt{\s})\,
 (\theta-\theta_0)\right\}\label{e1-e0}
\ee
where the subscript $0$ clearly refers to the background solution
$g_0$. In other words, the background terms can be considered  as
resulting from  the   embedding of  $(B_t,\s_{\mu\nu})$ into  a
reference spacetime and by requiring that the surface evolution is
the same in
$M$ and in the reference spacetime, i.e.  by requiring that the
boundary lapses $\bar N$ and $\bar N_0$ coincide, the boundary shifts
$\bar N^\al$ and $\bar N_0^\al$ are equal  as well as  the
time rates  of change
$\pounds_\xi(\sqrt{\s})$ of the  metric $\s$ in the two spacetimes
are the same;  see
\cite{Booth}. Roughly speaking, expression \form{e1-e0} defines the
quasilocal energy  of the solution $g$ \emph{relative} to  a solution
$g_0$  satisfying the same Dirichlet conditions on the boundary.
From a symplectic viewpoint we are selecting the six
components of the boundary metric
$\bar \g_{\mu\nu}$  as the control parameters of the gravitational
system, while the response variables  are the relative quasilocal
surface  energy $\bar \epsilon-\bar \epsilon_0$,   the relative
quasilocal surface  angular momentum $\bar j_\al-\bar j_{0\al}$ and
the relative  velocity parameter $\theta-\theta_0$.

We
stress that formula
\form{e1-e0} differs from the expression of quasilocal internal energy
as given in \cite{Booth}  only for the addition of the term
$\pounds_\xi(\sqrt{\s})\,
 (\theta-\theta_0)$.
[Indeed, in \cite{Booth}  the term $\theta\,\pounds_\xi(\sqrt{\s})$
was considered as a $p\,\dot q$ boundary term and thereby it was not
included  in the definition itself of the Hamiltonian].
 Hence formula
\form{e1-e0} perfectly agrees with the expression given in those
papers  when the control parameter $\sqrt{\s}$ is constant in time,
i.e. $\pounds_\xi(\sqrt{\s})=0$  (or when the intersection parameter
$\theta$ equals the intersection parameter $\theta_0$ in the
reference spacetime). Otherwise expression \form{e1-e0} contains an
additional contribution to the energy coming from the accretion in
time of the surface. However, we stress  that the quasilocal energy
contained in
$B_t$ depends not only on $B_t$ itself but also on its history,
geometrically described by the world tube $\B$. Notice that we can
always make use of  the condition $\pounds_\xi(\sqrt{\s})=0$ to assign
a world tube $\B$ to $B_t$ or, equivalently, to assign an evolution
vector field $\xi$ to $B_t$; see \cite{Kij}. In practice, however,
one always  assumes that the evolution vector field  is a Killing
vector field on the boundary. Indeed, this requirement guarantees 
the conservaton of energy in time, where the time  is defined by the
flow of the vector field. Provided that these remarks are taken
into account, the term $\pounds_\xi(\sqrt{\s})\,
 (\theta-\theta_0)$ can be set equal to zero  so that formula
\form{e1-e0} agrees with the one calculated in \cite{Booth}.

Another expression for the  energy of the gravitational field in the
case of a non--orthogonal foliation of spacetime was previuosly
calculated in \cite{HHnon}. The calculation there developed was
carried out with respect to the foliation of spacetime as a whole
rather then with respect to the foliation of the boundary, as instead
explicitly done in this Section. 

The expression there
presented, rewritten according to our
notation, reads as follows:
\ba
E_{HH}(g)-E_{HH}(g_0)&=&{1\over \kappa} \int_{B_t} d^2x\sqrt{\s}\left[
N {\cal K} -n_\mu N_\nu \left( {2\kappa\over \sqrt{h}
}P^{\mu\nu}\right)\right]\nonumber\\
&&+{1\over \kappa}\int_{B_t} d^2x\sqrt{\s} N \bar u^\mu\nabla_\mu
\theta \nonumber
\ea
The background solution $g_0$ was then chosen in  \cite{HHnon} by
requiring  that
$g$ and
$g_0$  induce the same  $3$--metric  on the boundary $\B$  and the
same intersection angle $\theta$ on $B_t$. Hence:
\ba
E_{HH}(g)&=&{1\over \kappa} \int_{B_t} d^2x\sqrt{\s}N\left[  {\cal
K}-{\cal K}_0\right]\label{HH}\\
&&-2\int_{B_t} d^2x\sqrt{\s}\left[n_\mu N_\nu \left( {P^{\mu\nu}\over 
\sqrt{h}
}\right)-n_{0\mu} N_{0\nu} \left( {P_0^{\mu\nu}\over \sqrt{h_0}
}\right)\right]
\nonumber
\ea
Taking into account the relation
\be
N_\al=h^\nu_\al\, N_\nu=(\s^\nu_\al +n^\nu\, n_\al)
N_\nu\um^{\form{vv}}\s^\nu_\al
\, N_\nu+N\, v\, n_\al
\ee
together with the splitting (see \cite{BLY})
\be
K_{\mu\nu}=l_{\mu\nu} +(u\cdot b)n_\mu \, n_\nu+ 2\, \s^\al_{(\mu}\,
n_{\nu)} \, K_{\al\b} \, n^\b\quad ( b^\mu=n^\nu\nabla_\nu n^\mu)
\ee 
(where $l_{\mu\nu}=
-\s^\al_\mu \s^\b_\nu\nabla_\al u_\b$ and $b^\mu=n^\nu\, \nabla_\nu\,
n^\mu$), expression \form{HH} can be further rewritten as:
\be
E_{HH}(g)=\int_{B_t}d^2x \,\sqrt{\s}\left\{
 {N\over\kappa}\left[({\cal K}-{\cal K}_0) + v\, (l-l_0)\right]
-N^\b \, \s^\al_\b (j_\al-j_{0\al})\right\}\label{HHH}
\ee
(with $l=\s^{\mu\nu}\, l_{\mu\nu}$ and $l_0=\s_0^{\mu\nu}\,
l_{0\mu\nu}=\s^{\mu\nu}\, l_{0\mu\nu}$).

In order to establish  a correspondence
between expression \form{e1-e0} and the definition \form{HHH},  we
first have to rewrite \form{e1-e0} in terms of ``unboosted''
quantities.  To reach this goal we shall make use  of the boost
relations (see again
\cite{BLY}):
\ba
&&\bar N=N/\g\\
&&\bar {\cal K}=\g \, {\cal K} +\g \,v\, l\\
&&\bar j_\al=j_\al -{1\over \kappa} \nabla_\al \theta
\ea
(where  $j_\al={1\over \kappa}\, \s_\al^\b\, K_{\b\mu}\,
n^\mu=-{2\over\sqrt{h}}
\s_{\al\mu}P^{\mu\nu} n_\nu$)
together with the relation 
\be
\pounds_\xi( \sqrt{\s})=\sqrt{\s}\left[ -N \,l -N\, v\, {\cal K}
+{\mathfrak D}_\al (\s^\al_\b N^\b)\right]
\ee
where ${\mathfrak D}_\al$  denotes the (metric) covariant derivative
on the surface. After a straightforward  calculation it is easy to
show that  expression
\form{e1-e0} exactly agrees with \form{HHH}
 provided the
same matching conditions are imposed between $g$ and  the background
solution
$g_0$.

For the sake of completeness we recall that a formula 
for the  energy of the gravitational field in the
case of a non--orthogonal foliation of spacetime can be also found
in \cite{Kij}. It was there obtained through a new method of
variation  of the gravitational Lagrangian and through 
a Legendre transformation tecnique. We end this section by noticing 
that   expressions
\form{Energy} and \form{kij} together reproduce the ``homogenous''
formula $(80)$ of Kijowski's paper \cite{Kij} which is at the basis of
his formalism.

In this way a (conditioned) correspondence among the various
definitions of energy  \cite{Booth,HHnon}, \cite{BY,BLY} and
\cite{Kij} (based, respectively, on  the canonical analysis, on
Hamilton--Jacobi analysis and on symplectic analysis)
and the  N\"other approach is established.

%%%%%%%%%%%%%%%%%%%%%%%%%%%%%%%%%%%%%%%%%%%%%%%%%%%%%%%%%%
\subsection{Entropy}
\label{entropy}

In the previous section we have shown that formula \form{Energy} leads
to the definition of quasilocal internal energy provided that
Dirichlet boundary conditions are imposed on the metric field.
Formula \form{Energy} can be also used in order to describe the
first law   of black hole thermodynamics. Indeed, the
direct evaluation of
\form{Energy} on the horizon of a black hole solution and the
boundary conditions  dictated by the geometry  of the black hole
horizon  lead to the formulation  of the first law of black hole
thermodynamics.

Indeed,   let us  specialize expression
\form{Energy} to an axisymmetric stationary black hole solution
$g_{\mu\nu}$ of Einstein (vacuum) field equations. 
On each leaf $\Sigma_t$ let us choose spatial coordinates
$x^i$ co--rotating with the horizon.  Hence the vector field
$\xi$  coincides  with the
null vector field  generating the horizon.
    We recall that, with this choice of coordinates, the metric near the
horizon  satisfies
 the regularity
condition:
\be
\bar n^\mu\partial_\mu N= \kappa_H\qquad (\kappa_H=\hbox{surface
gravity})
\label{regularity}
\ee
which ensues  from the necessity of removing the conical singularity 
in the complex metric  obtained from the Lorentzian metric by
analytically  continuing  time to imaginary values (see
\cite{BYdue},
\cite{CarterBH},
\cite{Brown}). We claim that 
formula
\form{Energy} already encloses  all the information about the first
law, in the same way it was
originally formulated in
\cite{BYdue}, but now generalized to non--orthogonal intersections
(see also \cite{BoothIH}). To realize this, first of all notice that
the integrand in the right hand side of formula
\form{Energy} is a closed $2$--form as $\xi$ is a
Killing vector field of the
solution $g$ (and hence  the right hand side of   \form{kij} is
vanishing). By Stokes' theorem, we have that
$\left.\delta_X
E(g)\right\vert_{B_t}=\left.\delta_X
E(g)\right\vert_{C_t}$ if $C_t$ is any $2$--surface
homologous  to
$B_t$, i.e. if $B_t-C_t$ is the
boundary of a $3$--dimensional region $\Sigma_t$. We can choose, for
example,
$C_t=\Delta$, where $\Delta$ denotes the (cross section of the)
black hole horizon. In that case the lapse and the shift are
vanishing on the horizon and, taking   the boundary 
condition \form{regularity} into account the only term surviving in
the right hand side of \form{Energy} is (see \cite{BoothIH}):
\be
\left.\delta_X
E(g)\right\vert_{\Delta}
={\kappa_H\over\kappa}\delta
\left(\int_H d^2x
\sqrt{\s}\right)={\kappa_H\over\kappa} \delta A_H
\label{aquarti}
\ee
where  $A_H$ denotes the area of the horizon. In order to attribute a
thermodynamical meaning  to the dynamics  of black holes, in
accordance with
\cite{BYdue} we define the
following intensive variables: the  \emph{inverse temperature} $\b$ of
the surface, the \emph{proper angular velocity} $\omega$
and  the \emph{surface pressure}
$p^{ij}$. They are defined,  respectively, as follows:
\ba
\b&=&{2\pi\over \kappa_H} \bar N\nonumber\\
\b \omega&=&{2\pi\over \kappa_H}\bar N^\phi\\
\b{\sqrt{\s}\over 2} p^{ij}&=&{2\pi\over \kappa_H}
\bar N{\sqrt{\s}\over 2} \bar s^{ij}\nonumber
\label{chemicalpotential}
\ea
As one can read off  from \form{Energy} they correspond to
$2\pi/\kappa_H$ times the  quantities conjugated, respectively,  to 
$\sqrt{\s}
\bar
\epsilon$, $\sqrt{\s}
\, \bar j_\al$  and $\s_{\al\b}$, i.\ e. conjugated to the extensive
variables (however, we stress that  the physical interpretation  of
quasilocal quantities  in terms of thermodynamical variables comes
from a path integral  formulation  of gravity -- see \cite{GH,BYdue}
-- and it is then  out of the scopes   of our discussion).
  Equating the left hand sides of
equations
\form{Energy} with
\form{aquarti} and taking  the property of stationarity  into account
we then obtain the first law of black hole thermodynamics for
spatially bounded system as  formulated in \cite{BYdue} but
generalized now to non--orthogonal intersections:
\be
\int_{B_t} d^2x\left\{ \b\delta (\bar \epsilon\,
\sqrt{\s})-\b\omega \delta(\bar j \sqrt{\s}) +\b {\sqrt{\s}\over 2}\,
p^{ij} \delta
\s_{ij}\right\}={\delta A_H\over 4}
\label{primoprinclocal}
\ee
(in geometric units $G=c=1$ where $\kappa=8\pi$). As already remarked 
in
\cite{BYdue}, the equation \form{primoprinclocal}  resembles  
the thermodynamical law
$T\delta S=
\delta E-\omega
\delta J +p
\delta V$  of thermodynamical systems provided we define $\b=1/T$ and
$S=A_H/ 4$ (and it reduces exactly to it in static, spherically
symmetric systems when $\b$, $\o$ and  $p^{ij}$ can be pulled out 
of the integral).

We point out  that the
equality in formula
\form{primoprinclocal}  is a trivial consequence of homological
properties, namely of Stokes' theorem. For
this reason  the formula can be generalized to solutions other than black
hole ones and, of course, it does not rely on horizons. Additional
boundaries (i.e. additional singularities other than a single black hole
one) are responsible of additional
contributions to the entropy. In this way, the homological nature of
entropy becomes evident; see
\cite{TaubBolt}, \cite{TaubNUTH}, \cite{TaubNUTHH}.

We also stress that formula \form{Energy}  is also well--suited to
describe  the thermodynamics of \emph{isolated horizons}; see
\cite{Ashtekar}.
 Indeed evaluating \form{Energy} on a isolated horizon  and taking
into account the  boundary conditions dictated by  the definition
itself of isolated horizons, the first principle  is recovered. We
shall analyse this matter in the  forthcoming paper \cite{HIO}.

%%%%%%%%%%%%%%%%%%%%%%%%%%%%%%%%%%%%%%%%%%%%%%%%%%%%%%%%%%%%%%%%
\section{Acknowledgments}
We are grateful  to G. Allemandi, L. Fatibene and M. Ferraris  of the
University of Torino for useful discussion on the subject. 
This work has been  partially supported  by the University of Torino
(Italy).
%%%%%%%%%%%%%%%%%%%%%%%%%%%%%%%%%%%%%%%%%%%%%%%%%%%%%%%%%%%%%%

%%%%%%%%%%%%%%%%%%%%%%%%%%%%%%%%%%%%%%%%%%%%%%%%%%%%%%%%%%%%%%%%%%%%%%%

\begin{thebibliography}{99}


\bibitem{BY}{J.\ D.\ Brown, J.\ W.\ York, Phys.\ Rev.\ D{\bf 47} (4),
1407 (1993).}



\bibitem{BYdue}{J.\ D.\ Brown, J.\ W.\ York, Phys.\ Rev.\ D
{\bf 47} (4),
1420 (1993). }

\bibitem{BTZRefB}{J.\ D.\ Brown, J.\ Creighton, R.\ B.\ Mann, Phys. Rev.
D{\bf 50}, 6394 (1994).}

\bibitem{Booth}{
I. Booth, gr-qc/0008030; I. Booth, R.B. Mann gr--qc/9810009,
gr--qc/9907077, gr--qc/9907072. }

\bibitem{BLY}{
J. D. Brown, S.R. Lau, J. W. York, gr-qc/0010024.}



\bibitem{Hayward} G. Hayward, Phys. Rev. D47, 3275 (1993).

\bibitem{Gravitation} {C.W. Misner, K.S. Thorne, J.A. Wheeler,
{\it Gravitation}, (Freeman, San Francisco,  1973)}

\bibitem{HHnon} 
S. W. Hawking, C. J. Hunter, Class. Quantum Grav.,
$\mathbf{13}$, $2735$, ($1996$).

\bibitem{Kij}{ J. Kijowski, Gen. Relativ. Gravit {\bf 29}, 307 (1997).
}

\bibitem{Silva} B. Julia, S. Silva, gr--qc/9804029;
 B. Julia, S. Silva, gr--qc/0005127.





\bibitem{Wald}{V. Iyer and R. Wald, Phys. Rev. D {\bf 50}, 846 (1994);
R.M.\ Wald, J.\ Math.\ Phys., {\bf 31},  2378 (1993).}


\bibitem{Wald95} V. Iyer, R. M. Wald, gr--qc/9503052.

\bibitem{BYOur}{
L.\ Fatibene, M.\ Ferraris, M.\ Francaviglia, M.\ Raiteri,
gr--qc/00030019, J.  Math. Phys., {\bf 42}, No. 3, 1173
 (2001). }





\bibitem{Remarks}{
L.\ Fatibene, M.\ Ferraris, M.\ Francaviglia, M.\ Raiteri,
Annals of Phys.
{\bf 275},  27 (1999).
}

\bibitem{Nester} C.-M. Chen,J. M. Nester, gr--qc/0001088.

\bibitem{PetrovKatz}{
J.\ Katz, D.\ Lerer, gr-qc/$9612025$,
J.\ Katz, J.\ Bicak, D.\ Lynden--Bell,  Phys.\ Rev.\ D{\bf 55} (10),
 5957 (1997),
A.\ N.\ Petrov, J.\ Katz,  gr-qc/$9905088$.
}


\bibitem{Cavalese}{M. Ferraris and M. Francaviglia, in: {\it 8th Italian
Conference on General Relativity and Gravitational Physics}, Cavalese
(Trento), August 30 --
September 3, World Scientific, (Singapore, 1988) 183; M. Ferraris and M.
Francaviglia, Gen. Rel.
Grav.  {\bf 22}
(9), 965 (1990). }

\bibitem{Robutti}{M.\ Ferraris, M.\ Francaviglia and O.\ Robutti, in:{\it
G\'eom\'etrie et Physique},
Proceedings of the {\it Journ\'ees Relativistes 1985} (Marseille, 1985),
Y.\ Choquet--Bruhat, B.\ Coll, R.\ Kerner, A.\ Lichnerowicz eds. Hermann,
(Paris, 1987) 112.}

\bibitem{Lagrange}{M. Ferraris, M. Francaviglia,
in: {\it Mechanics, Analysis and Geometry: 200 Years after Lagrange},
Editor: M. Francaviglia, Elsevier Science Publishers B.V., (Amsterdam,
1991) 451.}


\bibitem{RT}{T.\ Regge, C.\ Teitelboim, Annals of Physics {\bf 88},
286  (1974).}

\bibitem{CADM}{M.\ Ferraris and M.\ Francaviglia, Atti Sem. Mat. Univ. Modena,
{\bf 37}, , 61 (1989);
M.\ Ferraris and M.\ Francaviglia, Gen.\ Rel.\ Grav., {\bf 22}, (9),
965 (1990). }


\bibitem{MM} K.B. Marathe, G. Martucci, 
{\em  The Mathematical Foundations of Gauge Theories}, North Holland, 
(Amsterdam 
$1992$).

 \bibitem{Kolar} I. Kol\'a\v r, P.W. Michor, J. Slov\'ak, {\em Natural 
Operations in Differential Geometry},  Springer--Verlag,
  (New York, $1993$).




\bibitem{Trautman} {A. Trautman, in: {\it Gravitation: An Introduction to
Current Research}, L.
Witten ed.
(Wiley, New York, 1962) 168; A. Trautman,
     Commun. Math. Phys., {\bf  6},  248 (1967).}

\bibitem{Saunders} D.L. J. Saunders, {\em  The Geometry of Jet 
Bundles},  
Cambridge University Press (Cambridge, $1989$).

\bibitem{Sarda} G. Sardanashvily, {\em Generalized Hamiltonian 
Formalism for Field Theory}, World
Scientific (Singapore, $1995$).
  




\bibitem{FPC} M. Ferraris, in:  {\em Proceedings of the Conference on 
Differential Geometry and its
Applications}, Part $2$, Geometrical Methods in Physics, edited by D. 
Krupka (Brno,
Czechoslovakia, $1984$), pp. $61-91$. 

\bibitem{Komar} A.\ Komar, Phys. Rev. {\bf 113},  934 (1959).

\bibitem{Katz} J. Katz, Class.  Quantum Grav., $\mathbf{ 2}$, $ 
423$, ($1985$).

\bibitem{BTZ}{
L.\ Fatibene, M.\ Ferraris, M.\ Francaviglia, M.\ Raiteri,  Phys.
Rev. D{\bf 60}, 124012 (1999),
L.\ Fatibene, M.\ Ferraris, M.\ Francaviglia, M.\ Raiteri,  Phys.
Rev. D{\bf 60}, 124013 (1999).
}


\bibitem{TaubBolt}{ L.  Fatibene, M.  Ferraris, M.  Francaviglia, M.
Raiteri, Annals of Phys. {\bf 284}, 197 (2000),  gr-qc/$9906114$.}

\bibitem{Waldsymp} G. Burnett, R.M. Wald,  Proc. Roy. Soc. Lond.,
A{\bf 430}, 56 (1990); J. Lee, R.M. Wald, J. Math. Phys {\bf 31}, 725
(1990).

\bibitem{Sinicco}{M.\ Ferraris, M.\ Francaviglia and I.\ Sinicco, Il Nuovo
Cimento, {\bf 107B},
(11), 1303 (1992).}

\bibitem{forth} M.\ Francaviglia, M.\ Raiteri, \emph{Energy and
Boundary Conditions in Two Dimensional Gravity}, in preparation.




\bibitem{GH}{G.\ W.\ Gibbons, S.\ W.\ Hawking, Phys.\ Rev.\ 
D{\bf 15} (10),
2752(1977).}






\bibitem{CarterBH}{ B.\ Carter, in {\it Black Holes}, S. W. Hawking
and W. Israel editors (Cambridge
University Press, 1979).}


\bibitem{Brown}{ J.\  D.\  Brown, gr-qc/9506085.}



\bibitem{TaubNUTH}{C.\
J.\ Hunter, hep-th/9807010.}

\bibitem{TaubNUTHH}{S.\ W.\ Hawking, C.\ J.\ Hunter, hep-th/9808085.}



\bibitem{BoothIH} I.Booth, gr--qc/0105009.

\bibitem{Ashtekar} A. Ashtekar, C. Beetle,  S. Fairhurst, Class.
Quantum Grav.  {\bf 16}, L1 (1999);
A. Ashtekar, C. Beetle,  S. Fairhurst, Class.
Quantum Grav.  {\bf 17}, 253 (2000);
A. Ashtekar,  S. Fairhurst, B. Krishnan, gr--qc/0005083.


\bibitem{HIO} G.\ Allemandi, M.\ Francaviglia, M.\ Raiteri, \emph{The
First Law of Isolated Horizon via N\"other Theorem}, in preparation.







\end{thebibliography}
\end{document}